\documentclass[12pt]{article}
\usepackage{amsmath}
\usepackage{multirow}
\usepackage{amsfonts}
\usepackage{amssymb,graphics,psfrag}
\usepackage{array,epsfig,multirow,stmaryrd,graphicx}
\usepackage{comment}

\usepackage{slashed}

\usepackage{hyperref}

\def\hybrid{\topmargin -20pt    \oddsidemargin 0pt
        \headheight 0pt \headsep 0pt 
        \textwidth 6.25in      
        \textheight 9 in      
        \marginparwidth .875in
        \parskip 5pt plus 1pt
          \jot = 1.5ex
  }
\hybrid
\numberwithin{equation}{section}
\numberwithin{table}{section}

\newcommand{\beq}{\begin{equation}}
\newcommand{\eeq}{\end{equation}}
\newcommand{\be}{\begin{equation}}
\newcommand{\ee}{\end{equation}}
\newcommand{\bea}{\begin{eqnarray}}
\newcommand{\eea}{\end{eqnarray}}
\newcommand{\ben}{\begin{eqnarray*}}
\newcommand{\een}{\end{eqnarray*}}               
\newcommand{\ba}{\begin{aligned}}
\newcommand{\ea}{\end{aligned}}
\newcommand{\bt}{\begin{tabular}}
\newcommand{\et}{\end{tabular}}
\newcommand{\bc}{\begin{center}}
\newcommand{\ec}{\end{center}}

\newcommand{\cO}{\mathcal{O}}

\newcommand{\cD}{\mathcal{D}}
\newcommand{\cL}{\mathcal{L}}

\newcommand{\cK}{\mathcal{K}}

\newcommand{\nn}{\nonumber}

\newcommand{\cref}{{\bf [check ref]}}

\newcommand{\tr}{\mathrm{tr}\, }

\psfrag{n1}{$\nu_1$}
\psfrag{n2}{$\nu_2$}
\psfrag{n1'}{$\nu_1'$}
\psfrag{n2'}{$\nu_2'$}
\psfrag{n9}{$\nu_9$}
\psfrag{n10'}{$\nu_{10}'$}
\psfrag{t1}{${\nu}_1$}
\psfrag{t2}{${\nu}_2$}
\psfrag{t9}{${\nu}_9$}
\psfrag{t1'}{${\nu}_1'$}
\psfrag{t2'}{${\nu}_2'$}
\psfrag{t10'}{${\nu}_{10}'$}

\newcommand{\vecA}{A}
\newcommand{\rad}{r}

\usepackage{caption}
\usepackage{subcaption}

\def\blfootnote{\xdef\@thefnmark{}\@footnotetext}
\long\def\symbolfootnote[#1]#2{\begingroup%
\def\thefootnote{\fnsymbol{footnote}}\footnote[#1]{#2}\endgroup}

\begin{document}

\baselineskip=15pt

\begin{titlepage}
\begin{flushright}
\parbox[t]{1.8in}{\begin{flushright} MPP-2013-3\\CERN-PH-TH/2013-022\end{flushright}}
\end{flushright}

\begin{center}

\vspace*{ 1.2cm}

{\large \bf One-loop Chern-Simons terms in five dimensions}

\vskip 1.2cm

\begin{center}
 {Federico Bonetti\footnote{bonetti@mpp.mpg.de}, Thomas W.~Grimm\footnote{grimm@mpp.mpg.de} and Stefan Hohenegger\footnote{stefan.hohenegger@cern.ch}}
\end{center}
\vskip .2cm
\renewcommand{\thefootnote}{\arabic{footnote}}

{\footnotemark[1]\,\footnotemark[2] Max-Planck-Institut f\"ur Physik, \\
F\"ohringer Ring 6, 80805 Munich, Germany} 

{\footnotemark[3] Department of Physics, CERN - Theory Division, \\
 CH-1211 Geneva 23, Switzerland} 

 \vspace*{1cm}

\end{center}

\vskip 0.2cm
 
\begin{center} {\bf ABSTRACT } \end{center}
We compute one-loop corrections to five-dimensional gauge and gravitational Chern-Simons terms 
induced by integrating out charged massive fields. The considered massive fields are spin-1/2 and spin-3/2 fermions, as well as complex two-forms with first order kinetic terms. 
Consistency with six-dimensional gravitational anomalies of $(1,0)$ and $(2,0)$
theories is shown by interpreting the massive fields as excited Kaluza-Klein modes 
in a circle compactification. The results are in accordance with the geometric 
predictions of the M-theory to F-theory duality 
as well as the comparison with an explicit one-loop computation in heterotic string theory compactified on $K3\times S^1$. 

\hfill {February, 2013}
\end{titlepage}

\tableofcontents




\section{Introduction}

The derivation of a Wilsonian low-energy  effective action 
amounts to integrating out all excitations beyond a chosen cutoff mass scale and obtaining a  theory with modified couplings
for the remaining degrees of freedom. The consideration of such theories 
is crucial, for example, to study physical properties 
of a fundamental theory such as string theory at low energies. 
The corrections to the low energy effective action obtained by integrating out 
massive fields are organised in an expansion in the inverse  
mass scale. In the limit of large cutoff scale corrections are typically strongly suppressed and can be neglected. 
In this case all modes with masses above the cutoff scale become effectively non-dynamical and can be decoupled from the theory. 
This is the subject of 
well known results in quantum field theory,
such as the Appelquist-Carazzone-Symanzik
decoupling theorem \cite{Symanzik:1973vg}.  
This reasoning, however, breaks down for certain types of couplings.
Four-dimensional examples are furnished by  Goldstone-Wilczek currents \cite{Goldstone:1981kk}
and  Wess-Zumino terms \cite{D'Hoker:1984ph}
generated by integrating out a fermion
that becomes massive via Yukawa coupling to a scalar that gets a non-vanishing VEV.
They are  independent 
of the fermion mass and   have to be included 
in the low-energy effective action even in the limit 
in which it is taken to infinity.  
In this work we will study  couplings with similar features, namely gauge and gravitational Chern-Simons couplings in five-dimensional theories. 

The five-dimensional quantum field theories 
under consideration will propagate both massless 
and massive degrees of freedom. We will consider 
massive spin-1/2 fermions, spin-3/2 fermions, and 
complex two-forms. The kinetic and mass terms of 
the fermions are of standard form while the 
complex two-forms admit first order kinetic terms. The 
latter feature is possible in odd-dimensional theories 
and is crucial for the fields to introduce corrections 
to Chern-Simons couplings. This can be attributed 
to the fact that the Chern-Simons couplings violate 
parity and only fields with parity violating actions can modify their 
prefactors when deriving the Wilsonian effective action. 
The massive 
fields are minimally coupled to a massless $U(1)$ gauge 
field $A$ with field strength $F$. 
We aim to derive the corrections to the 
gauge Chern-Simons term $A\wedge F \wedge F$ and the 
gravitational Chern-Simons term $A \wedge \tr (R \wedge R) $, where $R$
is the five-dimensional curvature two-form, induced by 
integrating out all massive fields. 
After appropriate overall 
normalisation each of the massive fields yields an 
integer contribution to the Chern-Simons couplings. This
is consistent with the topological nature of the Chern-Simons couplings 
that implies that their prefactors are quantised and turn out to be 
independent of the mass scale of the fields 
that are integrated out. As a consequence, they survive 
the limit in which the mass scale is taken to infinity.
The results for the gauge Chern-Simons
coupling were given in \cite{Bonetti:2012fn}
without proof. 
This work substantiates 
our claims and extends them to include
gravitational couplings.
Our findings are 
 summarised in section \ref{Summary}.

This analysis is not purely academic, since, remarkably, these 
couplings elegantly encode information about higher-dimensional anomalies 
after  Kaluza-Klein compactification. In particular, we can consider a 
six-dimensional theory compactified on a circle to get a five-dimensional 
quantum field theory which propagates both massless and massive degrees of freedom. 
Six-dimensional gravitational anomalies
are thus associated to the five-dimensional
Chern-Simons couplings induced
by integrating out excited Kaluza-Klein modes.
Gauge anomalies can be accessed in the five-dimensional setup as well. In this case
the massive degrees of freedom can also arise from spontaneous gauge symmetry breaking.

As already explained in \cite{Bonetti:2012fn} (see also \cite{Cvetic:2012xn}), a heuristic explanation for the above mentioned connection between five-dimensional Chern-Simons terms and six-dimensional anomalies can be obtained by considering the one-loop diagram necessary to determine the latter. Indeed, in six dimensions, the gravitational anomaly is captured by a four-point amplitude with external graviton legs. The polarisations of the latter have to be contracted in all possible ways and thus particularly also include contributions corresponding to the compact $S^1$ direction. From the five-dimensional perspective, the four-point function therefore decomposes into a sum of correlators involving (five-dimensional) gravitons, the Kaluza-Klein vector $A$, and the graviscalar, which we will consider to be non-dynamical and replace by the radius $r$. Focusing on terms which break five-dimensional parity, we are naturally lead to consider Chern-Simons terms of the form $A\wedge F \wedge F$ and $A \wedge \tr (R \wedge R) $, where $R$
is the five-dimensional curvature two-form and $F$ the field-strength tensor corresponding to $A$. One-loop corrections to these couplings are therefore expected to encode information about higher-dimensional anomalies. We will have more to say about this interesting connection in the upcoming paper \cite{BGH}.

There are various ways to embed our five-dimensional or six-dimensional 
setup into string theory and M-theory. We will consider two realisations
in this work. Firstly, we will realise the five-dimensional setup 
by compactification of M-theory on an elliptically fibered 
Calabi-Yau threefold. Using the M-theory to F-theory limit reviewed in \cite{Denef:2008wq} 
the resulting low-energy effective action for the massless fields 
should be identified with the low energy effective action of a six-dimensional 
F-theory compactification on a circle after all Kaluza-Klein modes are 
integrated out. The Chern-Simons couplings on the M-theory 
side are determined by the intersection numbers and the second Chern class 
of the Calabi-Yau threefold \cite{Cadavid:1995bk, Antoniadis:1997eg, Bonetti:2011mw}. We thus find a purely geometric computation 
of the total Chern-Simons contribution for $(1,0)$ theories on a circle and 
agreement with our field theory computation can be shown. 

A second string theory realisation of our setup can be found by considering 
heterotic string theory on $K3\times S^1$ in the absence of NS5-branes (see e.g. \cite{Antoniadis:1995vz} for earlier computations in this setting). In this case the underlying six-dimensional $(1,0)$ theory is a theory with one 
self-dual and one anti-self dual tensor. The contributions to the  
five-dimensional Chern-Simons terms can in this case be computed as a three-point amplitude at one-loop. It turns out that the computation is largely insensitive to most of the details of the internal world-sheet CFT which allows us to calculate the amplitudes explicitly for generic points of the moduli space of $K3$ compactifications. The summation over the various massive modes propagating through the loop in the field theory picture corresponds to integrating over the moduli space of world-sheet tori from the string perspective that captures the contribution of all CFT excitations. We manage to perform this integration explicitly and find again agreement with our field theory result.

The paper is organised as follows. 
Section \ref{Summary} summarises the main results of the paper.
The families of massive fields that can generate one-loop
Chern-Simons terms are listed  in table \ref{summary_reps},
while table \ref{summary_table} gives 
the Chern-Simons coefficient for each of them.
Section \ref{field_theory_section} contains 
the Feynman diagram
computation of these coefficients. 
Section \ref{sec:Mtheory_Ftheory} discusses  the  check of the
field theory results in the framework of six-dimensional
F-theory compactifications. Section \ref{sec:heterotic} 
is devoted to the   explicit 
one-loop computation of the relevant amplitudes in
heterotic string theory on $K3 \times S^1$.
Finally, in the conclusions
we recapitulate our results
and discuss briefly further directions. 
The main body of the paper is accompanied 
by several appendices. Notations, conventions,
and other useful identities are collected in 
appendices  \ref{app_notations} and  \ref{app_pert_gravity}. The complete
Feynman rules used in section \ref{field_theory_section}
are gathered in appendix \ref{app_Feynman}.
In appendix \ref{App:Torus} we perform the 
 calculation of a torus integral that
appears in the string theory computation.


 \section{Summary of results} \label{Summary}

Let us start by summarising the results of this paper. The object of our investigation are five-dimensional
theories in which some massive fields
are coupled to a $U(1)$ gauge field $\vecA_\mu$ and to the metric $g_{\mu\nu}$.
In particular, we study how quantum corrections
due to massive fields can generate 
the
Chern-Simons couplings 
\beq \label{CS_couplings}
S_{AFF} = k_{AFF} \int \vecA \wedge F \wedge F  \ , \qquad
S_{ARR} =  k_{ARR} \int \vecA \wedge \tr(  R \wedge R) \ 
\eeq
in the low energy effective action. In these expressions
$F= d\vecA$ is the field strength of the $U(1)$ gauge field and $R$ denotes
the curvature two-form built from the metric $g_{\mu\nu}$.

We show that three classes of massive fields 
are capable of generating
such Chern-Simons terms in the 
quantum effective action:
massive spin-1/2  fermions $\psi$,
massive self-dual tensors $B_{\mu\nu}$, and 
massive spin-3/2  fermions $\psi_\mu$. 
By massive self-dual tensor we mean a complex two-form $B_{\mu\nu}$
that admits a non-standard first order kinetic term
$\bar B \wedge dB$ together with a mass term $m \bar B \wedge *B$.
Its free equation of motion thus reads schematically
\beq
* \! dB \propto m B \ .
\eeq
These tensor fields 
and their coupling to a 
$U(1)$ gauge field has been analysed in \cite{Townsend:1983xs, Bonetti:2012fn}. Further
details about massive self-dual tensors are given in section \ref{std_actions}.
We refer to these fields as self-dual because they can be thought of as
the excited Kaluza-Klein modes of a six-dimensional self-dual tensor
compactified on a circle.

 Spin-1/2 fermions,  self-dual tensors, and spin-3/2 fermions
can be characterised in terms of associated representations 
of the massive little group in five dimensions, $SO(4) \cong SU(2) \times SU(2)$.
Such representations  are labelled by a pair of half-integer 
spins $(j_1, j_2)$. The correspondence between massive fields and $SO(4)$ representations 
is summarised in table~\ref{summary_reps}. 
\begin{table}[!h]
\centering
\begin{tabular}{|  ccc   |} 
\hline 
 \rule[-.3cm]{0cm}{.8cm} field & free EOM & $SO(4)$ rep.  \\
 \hline 
\rule[-.3cm]{0cm}{.8cm} spin-1/2 fermion $\psi$  & $(\slashed{\partial} - c_{1/2} m)\psi =0$   &
$\left( \tfrac 12,0 \right)$ or $\left( 0, \tfrac 12 \right)$ \\
\rule[-.3cm]{0cm}{.8cm} self-dual tensor $B_{\mu\nu}$ & $(* d  - i c_B m )B =0 $ & $(1,0)$ or $(0,1)$ \\
\rule[-.3cm]{0cm}{.8cm} \hspace{.3 cm} spin-3/2 fermion $\psi_\mu$ \hspace{.3 cm} &
 \hspace{.3 cm} $(\gamma^{\rho\mu\nu} \partial_\mu  + c_{3/2}  m\gamma^{\rho\nu})\psi_\nu = 0$ \hspace{.3 cm} & 
 \hspace{.3 cm} $\left( \tfrac 12 , 1 \right)$ or $\left( 1, \tfrac 1 2 \right)$ \hspace{.3 cm} \\
\hline
\end{tabular}
 \caption{Summary of massive representations considered in this work.} \label{summary_reps}
\end{table}

We have included the equation of motion 
that puts each field   on-shell in the absence of interactions. The coefficients
 $c_{1/2}$, $c_B$, $c_{3/2}$ can take the values $\pm 1$ and
determine which $SO(4)$ representation is realised.
Note that here and in the following $m$ denotes the mass of the 
physical one-particle states and is thus taken to be positive.
The pairs of representations $(j_1,j_2)$ and $(j_2,j_1)$ are
interchanged   
under parity. Correspondingly, 
these classes of fields break parity at tree level. 
From this point of view, 
the fact that couplings of the form \eqref{CS_couplings}
are generated in the effective action can be interpreted as 
a parity anomaly: quantum effects 
compensate for the parity violation originally induced by these families 
of massive fields, after they are integrated out.

The following table summarises our findings 
for the coefficients $k_{AFF}$, $k_{ARR}$
 of the induced Chern-Simons couplings in \eqref{CS_couplings}. 
 Coefficients
$c_{1/2}$, $c_B$, $c_{3/2}$ correspond to those in table \ref{summary_reps}.
The symbol $q$ denotes the
$U(1)$ charge of the massive fields. 
 \begin{table}[!h]
\centering
\begin{tabular}{|lccc|} 
\hline
 \rule[-.3cm]{0cm}{.8cm}&  \quad spin-1/2 fermion  $\psi$  \quad &\quad  self-dual tensor $B_{\mu\nu}$ \quad & \quad spin-3/2 fermion $\psi_\mu$\quad
   \\ \hline
\rule[-.3cm]{0cm}{1.1cm} $ k_{AFF} =   $   & $\displaystyle - \frac{1}{48 \pi^2}  \, q^3  \cdot c_{1/2}$ & $\displaystyle - \frac{1}{48 \pi^2}  \, q^3  \cdot (- 4\, c_{B})$ 
& $ \displaystyle - \frac{1}{48 \pi^2}  \, q^3  \cdot (5 \, c_{3/2})$ \\
\rule[-.6cm]{0cm}{1.6cm} $k_{ARR} =   $   &   $\displaystyle - \frac{1}{384 \pi^2} \, q \cdot c_{1/2}$ & $\displaystyle - \frac{1}{384 \pi^2} \, q \cdot (8 \,c_B)$ 
& $\displaystyle - \frac{1}{384 \pi^2} \, q \cdot (-19 \, c_{3/2})$ \\
\hline
\end{tabular}
 \caption{Summary of the one-loop contributions for various fields. } \label{summary_table}
\end{table} 
It is a dimensionless quantity and its normalisation is fixed by the minimal coupling
 prescription $  \partial_\mu \rightarrow \partial_\mu - i q \vecA_\mu $. 
 The derivation of these results is the subject of the upcoming sections.
Nonetheless, let us stress here two crucial aspects
of the computation.
Firstly, $k_{AFF}$ and $k_{ARR}$ are quantum corrected at one-loop only.
This is expected by arguments involving locality of the effective action 
and quantisation of the Chern-Simons couplings
\cite{Witten:1996qb} and is 
consistent with the interpretation in terms of parity anomalies in five dimensions.

Secondly,
our  results   are derived 
using a simple quadratic action for the massive fields, 
which only includes minimal coupling to the gauge field $\vecA_\mu$
and the metric $g_{\mu\nu}$. 
We argue that $k_{AFF}$ and $k_{ARR}$
are indeed insensitive to any fine detail of the 
interactions. 
For the $k_{AFF}$ coupling, the effect of some non-minimal interactions
is analysed explicitly in section \ref{sec:non_minimal}. It is shown there that such 
non-minimal couplings
do not affect the renormalised value of $k_{AFF}$.
These features are expected 
for topological couplings such as \eqref{CS_couplings}
that can be interpreted as parity anomalies.

Note that we refrain from a discussion about 
the possibility to write down 
fully consistent interacting theories for the 
three classes of massive fields under examination. For instance,
it is expected that an interacting theory of massive spin-3/2 fermions
is only possible in presence of  (possibly 
spontaneously broken) supersymmetry,
even though our findings are independent of the
precise way it is realised in the five-dimensional action.
From this point of view, 
we do not consider 
other parity-violating representations
of $SO(4)$, such as $(\tfrac 32,0)$ or $(2,0)$,
because no example is known of 
consistent interacting theories
for the corresponding massive fields.


 \section{Field theory computation} \label{field_theory_section}

In this section we compute the coefficients of the Chern-Simons
couplings \eqref{CS_couplings} in perturbative quantum field theory.
We start by reviewing the actions for the massive spin-1/2 fermion, self-dual tensor, 
and spin-3/2 fermions minimally coupled to 
the $U(1)$ gauge field and the metric. We then describe the main points of the Feynman diagram
calculations for the gauge and the gravitational Chern-Simons terms.
We conclude the section by studying the effect of some non-minimal couplings
on the gauge Chern-Simons term.

 \subsection{Minimally coupled massive actions} \label{std_actions}

The Chern-Simons couplings \eqref{CS_couplings}
can be captured by one-loop computations
in a theory where the massive fields considered above
are minimally coupled to the $U(1)$ gauge field $\vecA_\mu$
and the metric $g_{\mu\nu}$. 
In this section we briefly review the corresponding actions.

A spin-1/2 fermion is described by a five-dimensional Dirac spinor $\psi$. In order to couple it to
the metric $g_{\mu\nu}$ we have to introduce a vielbein ${e^a}_\mu$.
The action for $\psi$ minimally coupled to the $U(1)$ gauge field $\vecA_\mu$
and the vielbein ${e^a}_\mu$ is taken to be 
\beq \label{std_action_spin12}
S_{1/2}
= \int d^5 x \, e  \left[
 -  \bar\psi \gamma^\mu \cD_\mu \psi
 +  c_{1/2} m\,  \bar\psi \psi
  \right] \ ,  \quad c_{1/2}=\pm 1 \ ,
\eeq
where $e = {\rm det}\, {e^a}_\mu$, $\gamma^\mu = \gamma^a {e_a}^\mu$, and where
we have introduced the full spacetime and $U(1)$ covariant derivative
\beq \label{full_cov_der_spin12}
\cD_{\mu} \psi = \partial_\mu \psi + \tfrac 14 \omega_{\mu a b } \gamma^{ab}  \psi 
- i q \vecA_\mu \psi \ .
\eeq
On the right hand side, $\omega_{\mu ab}$ is the Levi-Civita spin connection 
constructed from the vielbein, and  $q$ is the $U(1)$ charge of the fermion $\psi$.
More details about our spacetime and gamma-matrix conventions can be found in appendix \ref{app_notations}.
As stated in section \ref{Summary}, $m$ is the positive physical mass
and $c_{1/2}$ labels two inequivalent spinor representations of the massive little group $SO(4)$ 
in five dimensions. Under a parity transformation, the sign of $c_{1/2}$ is reversed. 

Let us now turn to  massive self-dual tensors in five-dimensions. 
Their action, including
the coupling to a  $U(1)$ gauge field,   can be written as \cite{Bonetti:2012fn}
\beq \label{std_action_tensors}
S_{B}
=
\int d^5 x  \sqrt{-g} \left[
- \tfrac 14 i c_B\, \epsilon^{\mu\nu\rho\sigma\tau} \bar B_{\mu\nu} \cD_{\rho} B_{\sigma \tau}
- \tfrac 12  \, m \bar B_{\mu\nu} B^{\mu\nu} 
\right] \ , \quad c_B =\pm1 \ .
\eeq
The relevant part of the spacetime and $U(1)$ covariant derivative reads
\beq \label{full_cov_der_tensors}
\cD_{[\rho} B_{\mu\nu]} = \partial_{[\rho} B_{\mu\nu]} - i q \vecA_{[\rho} B_{\mu\nu]} \ .
\eeq
Note  that $g = {\rm det}\, g_{\mu\nu}$ and that $\epsilon^{\mu\nu\rho\sigma\tau}$ denotes the five-dimensional
Levi-Civita tensor. In our conventions, it satisfies $\epsilon^{01234} = -1/\sqrt{-g}$
if $0, \dots,4$ are curved indices.
Note that in this 
case parity violation is not due to the mass term, but to the kinetic term.

Finally, a spin-3/2 fermion is 
described by a Dirac vector-spinor  $\psi_\mu$ with action
\beq \label{std_action_spin32}
S_{3/2}
= \int d^5 x \, e  \left[
 -  \bar\psi_\rho \gamma^{\rho\mu\nu} \cD_\mu \psi_\nu
 - c_{3/2} m \, \bar\psi_\mu \gamma^{\mu\nu} \psi_\nu
  \right] \ , \quad c_{3/2} =\pm 1 \ ,
\eeq
where the antisymmetric part of the spacetime and $U(1)$ covariant derivative is given by
\beq \label{full_cov_der_spin32}
\cD_{[ \mu} \psi_{\nu ]} = \partial_{[\mu} \psi_{\nu]} + \tfrac 14 \omega_{[\mu| a b } \gamma^{ab}  \psi_{\nu]} 
- i q \vecA_{[\mu} \psi_{\nu]} \ .
\eeq
In analogy with the spin-1/2 case, 
the two inequivalent representations of $SO(4)$
differ by the sign of the mass term.

\subsection{Computation of the $\vecA \wedge F \wedge F$ coupling} \label{secAFF}

The  $U(1)$ Chern-Simons coupling $\vecA \wedge F \wedge F$
 does not involve the gravitational field. 
As a consequence, throughout this section
we can ignore the coupling of massive fields to gravity
and take $g_{\mu\nu} = \eta_{\mu\nu}$.
No distinction between flat and curved indices is made.
The coupling to $\vecA_\mu$ can be treated perturbatively
in the framework of quantum
field theory on flat spacetime.

The coefficient of the $\vecA\wedge F \wedge F$ term
in the quantum effective action
can be extracted from the three-point function
of the gauge field $\vecA_\mu$.
More precisely, we work in momentum space and 
we denote by $\Gamma_{AAA}$
the sum of 1PI Feynman diagrams with three
external vectors with incoming momenta $p_1$, $p_2$, $p_3$
and polarisation vectors $e_1, e_2, e_3$. The Chern-Simons term
\beq
k_{AFF} \int  \, \vecA \wedge F \wedge F =
- k_{AFF} \int d^5x \, \epsilon^{\mu\nu\rho\sigma\tau} 
\vecA_\mu \partial_\nu \vecA_\rho \partial_\sigma \vecA_\sigma
\eeq
in the effective action corresponds to a contribution
to $\Gamma_{AAA}$ of the form
\beq \label{relevant_AAA}
\Gamma_{AAA} \supset  i3! \times (- k_{AFF})\, \epsilon_{\lambda \tau \mu_1 \mu_2 \mu_3} \, 
p_1^\lambda \, p_2^\tau\, e_1^{\mu_1}  e_2^{\mu_2}  e_3^{\mu_3} \ ,
\eeq
where 
we have included a factor of $i$ from the Feynman rules
and the combinatorial factor $3!$ to take into account
symmetry under permutations of the three vectors.
Contributions to $\Gamma_{AAA}$ different from 
\eqref{relevant_AAA} will be ignored. They correspond to higher-derivative and 
non-local terms in the effective action. As already mentioned, we expect that the right hand side of \eqref{relevant_AAA}
is corrected at one loop only.
As shown in sections \ref{sec:Mtheory_Ftheory} and \ref{sec:heterotic}
our one-loop results pass non-trivial tests 
in the framework of F-theory and heterotic string theory.

We can derive Feynman 
rules using the actions (\ref{std_action_spin12}), (\ref{std_action_spin32}) and (\ref{std_action_tensors}) evaluated in flat spacetime 
and extract the propagators for massive fields, together with the 
interaction tri-vertex among two massive fields and one gauge 
field $\vecA_\mu$. These propagators and vertices are listed in appendix 
\ref{app_Feynman}.

\begin{figure}
\centering
\includegraphics[trim =130 500 250 70 ,clip, scale=0.7]{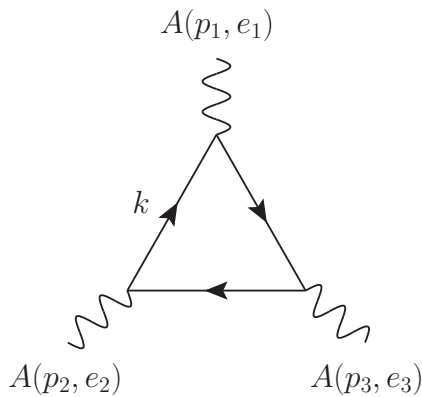}
\caption{One-loop Feynman diagram involved in the computation
of the Chern-Simons coefficient $k_{AFF}$. The external lines are three vectors $\vecA$
with incoming momenta $p_1$, $p_2$, $p_3$ and polarisation vectors
$e_1$, $e_2$, $e_3$. The internal lines can represent a
massive spin-1/2 fermion, a massive self-dual tensor,
or a massive spin-3/2 fermion. The loop momentum $k$
flows in the direction of the arrow.}
\label{diagram_AFF}
\end{figure}

At the one-loop level, only one class of diagrams
can be built using the interaction vertices at hand. A representative diagram
is depicted in figure \ref{diagram_AFF}. 
Wiggly lines represent the external vectors, 
while solid lines represent massive fields. Each class of massive fields
contributes separately to the amplitude.
To get the full answer,
it has to be summed with the analog diagram where the orientation of the loop
is reversed.
This is equivalent to 
swapping the labels 1 and 2 on the external legs.
Since the relevant structure in \eqref{relevant_AAA}
is invariant under this relabelling, 
the loop-reversed diagram simply gives an
overall  additional factor 2.

The denominator of the diagram (which is determined through its propagator factors) is the same for all
fields running in the loop. If the labelling of figure \ref{diagram_AFF} is adopted, it is 
given by
\begin{equation}
\mathbb{D} =  \frac{1}{k^2 + m^2} \frac{1}{(k-p_2)^2 + m^2} \frac{1}{(k + p_1)+m^2} \ ,\label{Denom}
\end{equation}
which is to be completed by a suitable numerator factor $\mathbb{N}$ which particularly encodes information about the vertices and is strongly dependent on the fields running in the loop. In (\ref{Denom}), the usual Feynman $i\epsilon$ prescription is understood.
We make use of Schwinger parameterisation to unify denominators, and write
\begin{equation}
\mathbb{D}= \frac{1}{m^6}
 \int_0^\infty d\alpha \int_0^\infty d\beta \int_0^\infty d\gamma \; e^{-(\alpha + \beta + \gamma)(\ell^2 + \Delta)/m^2} \ .
\end{equation}
In this expression, $\alpha, \beta, \gamma$ are dimensionless parameters, and we have made 
use of the shorthand notations
\begin{equation} \label{ell_and_Delta}
 \ell = k - y p_2 + z p_1 \ , \quad
 \Delta = m^2 + 2 y z p_1 \cdot p_2 + y(1-y) p_1^2 + z(1-z) p_2^2 \ ,
\end{equation}
where $y=\beta/(\alpha + \beta + \gamma)$ and $z = \gamma/(\alpha + \beta + \gamma)$.
The full diagram is then given by
\beq \label{dgrm}
\mathbb{I}=\mathbb{D}\cdot \mathbb{N}= \frac{1}{m^6}
 \int_0^\infty d\alpha \int_0^\infty d\beta \int_0^\infty d\gamma \int \frac{d^5 \ell}{(2\pi)^5}\; 
 e^{-(\alpha + \beta + \gamma)(\ell^2 + \Delta)/m^2} \, 
\mathbb{N}  \ ,
\eeq
where, of course, the numerator is different for 
different species of massive fields running in the loop. 
We also note that the sum of the diagram in figure \ref{diagram_AFF} 
 with the diagram with the opposite orientation
has
a distinct symmetry with respect to exchanging the 
external points. On general grounds, one can show that 
this symmetries restrict the parity violating part of the integrand 
in \eqref{dgrm} at the bilinear level in the external 
momenta to only depend on the Schwinger parameters 
in the combination $(\alpha + \beta + \gamma)$. 
This is a useful consistency check we have applied throughout the computations.

By naive power-counting arguments, we do
not expect any infrared divergence in this one-loop diagram,
but we cannot exclude the possibility of ultraviolet divergences. If Schwinger
parameterisation is used, the integral over the loop momentum $\ell$
contains an exponential factor and (after Wick rotation)
 is convergent as long as $\alpha + \beta + \gamma$
is strictly positive. Ultraviolet divergences are  translated into 
divergences
in the $\alpha,\beta,\gamma$ integration, coming from the region where 
these three parameters are simultaneously small. 
We regularise the amplitude by cutting out this portion of the $\alpha,\beta,\gamma$
integration domain with a step-function: in \eqref{dgrm}
we make the replacement
\beq \label{reg_prescription}
 \int_0^\infty d\alpha \int_0^\infty d\beta \int_0^\infty d\gamma \quad
 \rightarrow  \quad
 \int_0^\infty d\alpha \int_0^\infty d\beta \int_0^\infty d\gamma \;
 \theta(\alpha + \beta + \gamma - \epsilon) \ ,
\eeq
where $\epsilon>0$ is the regulator.

Recall from \eqref{relevant_AAA} that we are only interested in the coefficient
of a term with two powers of external momenta contracted with an $\epsilon$-symbol.
 This allows
us to simplify the computation of the diagram.

First of all, only the terms 
that contain an $\epsilon$-symbol have to be kept in the numerator. 
If a self-dual tensor runs in the loop, the $\epsilon$-symbol
is introduced directly at the level of Feynman rules both in the propagator
and in the vertex.
When a spinor runs in the loop, 
the $\epsilon$-symbol is generated 
by traces of gamma matrices.
This follows from the identities
\beq
{\rm tr} \, 1 = 4 \ , \quad
{\rm tr} \, \gamma^{\mu_1 \mu_2 \mu_3 \mu_4 \mu_5} = 
4 i \, \epsilon^{\mu_1 \mu_2 \mu_3 \mu_4 \mu_5} \ , \quad
{\rm tr} \, \gamma^{\mu_1 \dots \mu_p} = 0 \text{ for $p=1,2,3,4$}.
\eeq
We see that only those terms need to be retained that
contain an odd number of gamma matrices greater than or equal to five.

Second of all, 
we can  perform a formal 
power series expansion of \eqref{dgrm} in $p_1$, $p_2$ and 
we can neglect all terms that are not bilinear in $p_1$ and $p_2$.
In particular, this implies that we can use the  approximation
$\Delta \approx m^2$, 
since all other terms in the exact expression 
\eqref{ell_and_Delta} for
$\Delta$ would 
generate additional
powers of external momenta of the form $p_1^2$, $p_2^2$, or $p_1 \cdot p_2$.

Finally, by symmetry arguments (not spoiled
by our choice of regulator),
we can make the following replacements in the numerator under the $\int d^5 \ell$ integral:
\begin{align} \label{eq:ell_replacements}
 & \ell_{\mu_1} \dots \ell_{\mu_r} \rightarrow 0  \text {   if $r$ is odd}\ , \nn\\
 & \ell_\mu \ell_\nu \rightarrow \tfrac 15 \, \ell^2 \, \eta_{\mu\nu} \ , \quad  \quad
 \ell_{\mu_1} \ell_{\mu_2} \ell_{\mu_3} \ell_{\mu_4} \rightarrow \tfrac{1}{35} \, (\ell^2)^2
 (\eta_{\mu_1 \mu_2} \eta_{\mu_3 \mu_4} + \eta_{\mu_1 \mu_3} \eta_{\mu_2 \mu_4}) \ , \quad \quad \dots
\end{align}
All tensor integrals in the loop momentum are thus reduced to scalar integrals.

The calculation of the diagram is now straightforward
but tedious.\footnote{We made use of the {\it Mathematica} packages {\it xTensor} of
the bundle {\it xAct} \cite{xAct}
and {\it GAMMA} \cite{Gran:2001yh}.} After the numerator algebra is performed and the 
replacements \eqref{eq:ell_replacements} are made, 
the integrals over the loop momentum and the Schwinger
parameters are computed using the formulae
\begin{align}
& \int \frac{d^5 \ell}{(2\pi)^5} \, e^{-(\alpha + \beta +\gamma)\ell^2 /m^2}
(\ell^{2})^n = 
 \frac{i m^{2n+5}}{24 \pi^3} \frac{\Gamma(n+5/2)}{(\alpha + \beta + \gamma)^{n+5/2}} \ , \\
& \int_0^\infty d\alpha \int_0^\infty d\beta \int_0^\infty d\gamma \; \theta(\alpha + \beta + \gamma - \epsilon) \;
\frac{  e^{-(\alpha + \beta + \gamma)}  } {(\alpha + \beta + \gamma)^a} \, 
\alpha^{n_1} \beta^{n_2} \gamma^{n_3}
 =\nn  \\
 &\qquad \qquad \qquad \qquad = \frac{\Gamma(1+n_1) \Gamma(1+n_2) \Gamma(1+n_3)  }{\Gamma(3  + n_1 + n_2 + n_3)} \Gamma(3 + n_1 + n_2 + n_3 -a ; \epsilon) \ . \label{gamma_integral} 
\end{align}
We have performed
the usual Wick rotation $\ell^0 \rightarrow i \ell^0$ in the first integral and
have introduced the incomplete gamma function
\begin{equation}
 \Gamma(x;\epsilon) = \int_\epsilon^\infty d\tau \tau^{x-1} e^{-x} 
\end{equation}
in the second integral.

Let us consider the diagram where the spin-1/2 fermion $\psi$ runs in the loop.
By power-counting we expect a quadratic divergence, since the numerator 
has up to three powers of the loop momentum. The parity-violating part of the numerator,
however, turns out to be of zero-th order in the loop momentum,
thus giving a finite result without the need of any regulator.

This does not hold for the diagrams where 
 $B_{\mu\nu}$ and $\psi_\mu$ run in the loop.
In fact, even though the parity-violating part of the numerator 
has a better UV behaviour than
the full diagram, it still contains terms proportional to $\ell^2$
or $(\ell^2)^2$. This implies that both diagrams 
have a divergent piece. In our regularisation scheme
such divergences appear as coefficients 
of negative powers of the regulator $\epsilon$
in a formal expansion of the diagram.

We can then give the $\epsilon$-expansion 
for all the three species under consideration: 
spin-1/2 fermions $\psi$, tensors $B_{\mu\nu}$, and spin-3/2 fermions
$\psi_\mu$,
\begin{align} ({\rm diagram})_{1/2}   
&=  \frac{i}{64 \pi^2} \,c_{1/2} \, q^3 &
 &\negthickspace \negthickspace \negthickspace \negthickspace \negthickspace \negthickspace \bigg[ &
 & \negthickspace \negthickspace \negthickspace \negthickspace \negthickspace \negthickspace &
 & \negthickspace \negthickspace \negthickspace \negthickspace \negthickspace \negthickspace +4 &
 & \negthickspace \negthickspace  \negthickspace \negthickspace \negthickspace \negthickspace 
 + \cO(\epsilon^{1/2}) \bigg] \ , \label{finite_and_divergent_12}
\\  ({\rm diagram})_{B}  
& =  \frac{i}{64 \pi^2} \,  c_{B} \,\,\,\, q^3& 
&\negthickspace \negthickspace \negthickspace \negthickspace \negthickspace \negthickspace \bigg[ &
 & \negthickspace \negthickspace \negthickspace \negthickspace \negthickspace \negthickspace+\frac{15 }{\sqrt{\pi}}  \epsilon^{-1/2} &
 & \negthickspace \negthickspace \negthickspace \negthickspace \negthickspace \negthickspace-16 &
 & \negthickspace \negthickspace \negthickspace \negthickspace \negthickspace \negthickspace
 + \cO(\epsilon^{1/2}) \bigg] \ ,  \label{finite_and_divergent_B}
 \\
  ({\rm diagram})_{3/2}  
 & =   \frac{i}{64 \pi^2} \,c_{1/2} \, q^3&
&\negthickspace \negthickspace \negthickspace \negthickspace \negthickspace \negthickspace\bigg[   - \frac{105 }{4 \sqrt{\pi}} \epsilon^{-3/2}    &
 & \negthickspace \negthickspace \negthickspace \negthickspace \negthickspace \negthickspace+ \frac{15}{4 \sqrt{\pi}} \epsilon^{-1/2}  &
 & \negthickspace \negthickspace \negthickspace \negthickspace \negthickspace \negthickspace+20 &
 & \negthickspace \negthickspace \negthickspace \negthickspace \negthickspace \negthickspace
 + \cO(\epsilon^{1/2}) \bigg] \ . \label{finite_and_divergent_32}
\end{align}
Note that the factor $(-1)$ for a fermionic loop has been taken into account,
but we have not inserted the overall factor $2$ due to the diagram
with the reversed loop orientation.

In order to extract the physical observable $k_{AFF}$
from these expressions we adopt a minimal subtraction prescription:
negative powers of $\epsilon$
in the  expansion are discarded. This gives the results of table \ref{summary_table}.
In section \ref{sec:non_minimal} we 
discuss the effect of non-minimal
couplings and show how they can be used to
cancel divergences.

\subsection{Computation of the $\vecA \wedge \tr (R \wedge R)$ coupling} \label{secARR}

Let us  now turn to the discussion of the mixed $U(1)$-gravitational Chern-Simons term
$\vecA \wedge \tr (R \wedge R)$.
To compute its coefficient we
treat the coupling of massive fields 
to gravity perturbatively. The metric is 
written as 
\beq \label{metric_expansion}
g_{\mu\nu} = \eta_{\mu\nu} + h_{\mu\nu} \ ,
\eeq
and computations are performed order by order in a formal
power series in $h_{\mu\nu}$
around flat spacetime. Indices $\mu,\nu, \dots$ are thus 
raised and lowered
with $\eta_{\mu\nu}$ and its inverse and no distinction is made
between flat and curved indices.
Further details
about the expansion in $h_{\mu\nu}$
are collected in 
appendix \ref{app_pert_gravity}.

When $\vecA \wedge \tr( R \wedge R)$
is expanded according to \eqref{metric_expansion}
terms with arbitrarily high powers of $h_{\mu\nu}$
are generated, because of the non-linear
dependence of the Riemann tensor on the metric.
Nonetheless, in order to read off the Chern-Simons coupling
we can restrict to the lowest order term,
\begin{align} \label{ARR_expression}
&k_{ARR} \int \vecA \wedge \tr ( R \wedge R) =\\
&=- \tfrac 12 k_{ARR} \int d^5x \, \epsilon^{\mu_1 \mu_2 \mu_3 \mu_4 \mu_5}
A_{\mu_1}  \partial_{\lambda} \partial_{\mu_2} h_{\tau \mu_3} \left[ 
\partial^\tau \partial_{\mu_4} {h^\lambda}_{\mu_5} 
- \partial^\lambda \partial_{\mu_4} {h^\tau}_{\mu_5}  \right] + \cO(h^3)\nn \ .
\end{align}
As a consequence, the constant $k_{ARR}$
can be extracted from the sum of 1PI Feynman diagrams
with one vector and two gravitons, denoted $\Gamma_{Ahh}$.
More precisely, the sought-for Chern-Simons coupling
corresponds to the contribution
\beq \label{reading_kARR}
\Gamma_{Ahh} \supset i 2! \times \tfrac 12 k_{ARR} \,
 \epsilon_{\mu_0 \mu_1 \mu_2 \lambda \tau}\,  p_{1}^{ \lambda} p_{2}^{  \tau}
\left( p_{1\, \nu_2}  p_{2\, \nu_1} - \eta_{\nu_1 \nu_2} p_1 \cdot p_2 \right) 
e_0^{\mu_0} e_1^{\mu_1 \nu_1} e_2^{\mu_2 \nu_2} \ ,
\eeq
where $p_1$, $p_2$ are the incoming momenta of the gravitons,
$e_0$ is the polarisation tensor of the vector, 
and $e_1, e_2$ are the symmetric polarisation tensors of the gravitons.
The prefactor $i2!$ comes from the standard Feynman rule prescriptions.
Any term that does not match the structure of the right hand side of 
\eqref{reading_kARR} will be neglected, since it would
correspond to higher-derivative and non-local terms
in the effective action.

It is interesting to note that the tensor structure  in
\eqref{reading_kARR}
is transverse with respect to 
both the vector and the graviton polarisation tensors,
i.e.~it vanishes if any of the replacements
\beq
e_0^\mu \rightarrow p_0^\mu = -p_1^\mu - p_2^\mu \ , \quad
e_1^{\mu \nu} \rightarrow a^{(\mu} {p_1}^{\nu)} \ , \quad 
e_2^{\mu \nu} \rightarrow a^{(\mu} {p_2}^{\nu)} 
\eeq
is made, for arbitrary $a^\mu$.
It can be shown that this tensor structure is the only structure
with an $\epsilon$-symbol and four powers of external momenta
that has this transversality property and is 
symmetric
in the exchange of labels 1 and 2.
Its appearance is a consequence of gauge invariance.
Transversality with respect to $e_0$ reflects invariance of 
 \eqref{ARR_expression} under $U(1)$ transformations.
Transversality with respect to $e_1$, $e_2$
 derives from invariance of $\eqref{ARR_expression}$
 under diffeomorphisms. Recall that under 
 an infinitesimal diffeomorphism with parameter $\xi^\mu$ we have
\beq
\delta h_{\mu\nu} = 2\partial_{(\mu} \xi_{\nu)} + \cO(h) \ .
\eeq
Gauge invariance can be used as a self-consistency check
of the Feynman diagram computation. Indeed, we find that 
the desired contributions to $\Gamma_{Ahh}$ organise 
into the structure \eqref{reading_kARR} after all relevant
diagrams are summed.

The Feynman rules 
needed in the diagrammatic computation of $\Gamma_{Ahh}$ are deduced by 
expanding the actions \eqref{std_action_spin12}, \eqref{std_action_spin32},
\eqref{std_action_tensors} for the massive fields
according to  \eqref{metric_expansion}.
This gives interaction vertices of arbitrarily high powers in $h_{\mu\nu}$ 
but  we only need
an expansion up to second order in $h_{\mu\nu}$.
More precisely, four kinds of vertices 
are relevant for the calculation of $\Gamma_{Ahh}$.
If we denote any of the massive fields 
$\psi$, $B_{\mu\nu}$, $\psi_\mu$ as $\Phi$,
we need:
the gauge tri-vertex $\bar\Phi \Phi \vecA$, already considered in the previous section;
the gravitational tri-vertex $\bar\Phi \Phi h$;
the purely gravitational quadri-vertex $\bar\Phi \Phi h h $;
 the mixed gauge-gravitational quadri-vertex $\bar\Phi \Phi \vecA h$.
All such vertices are collected  in appendix \ref{app_Feynman}.

\begin{figure}
\centering
\begin{subfigure}[b]{0.4\textwidth}
\centering
\includegraphics[trim= 130 510 250 60 ,clip, scale=0.7]{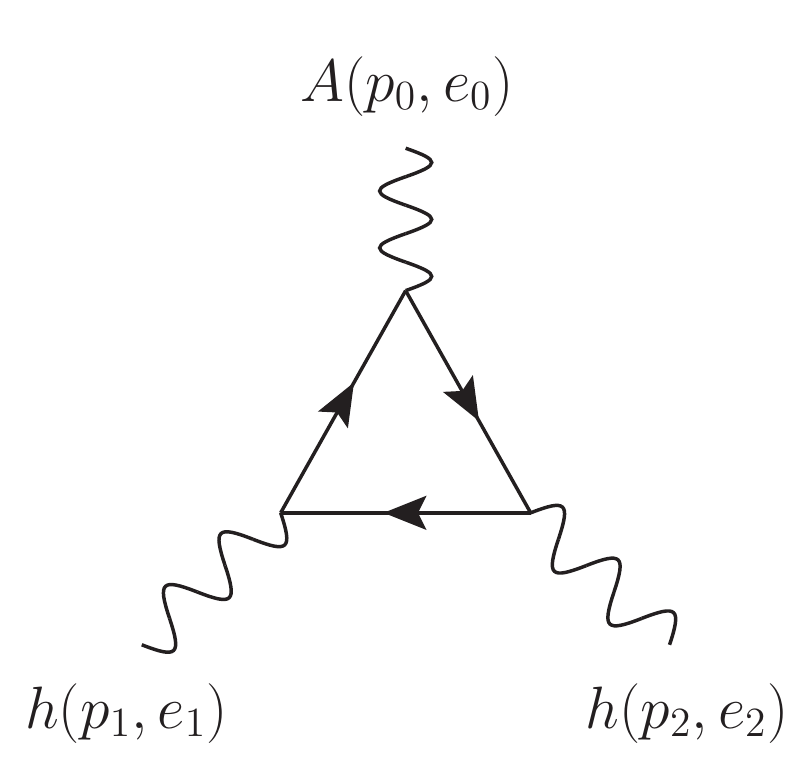}
\caption{}
\end{subfigure}
\quad \quad \quad \quad \quad \quad
\begin{subfigure}[b]{0.4\textwidth}
\centering
\includegraphics[trim = 180 500 260 70 ,clip, scale=0.7]{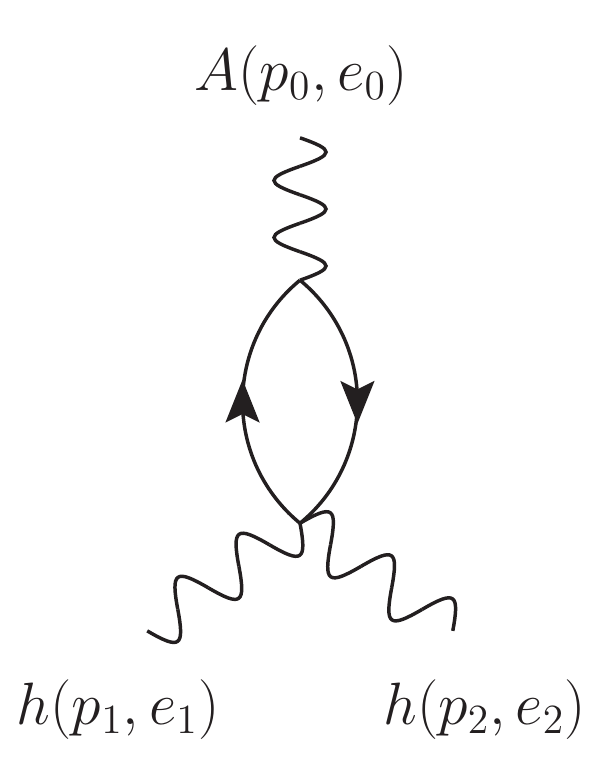}
\caption{}
\end{subfigure}
\\
\begin{subfigure}[b]{0.4\textwidth}
\centering
\includegraphics[trim= 170 500 230 110 , clip, scale=0.7]{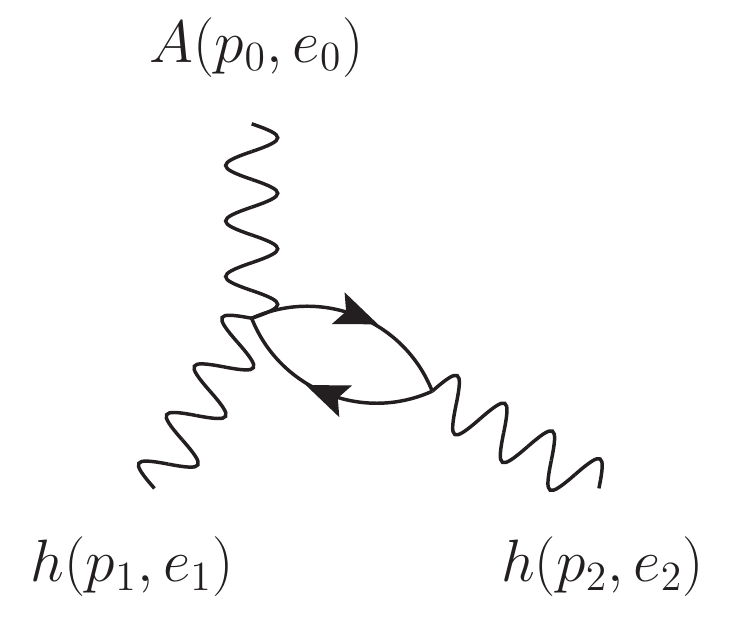}
\caption{}
\end{subfigure}
\caption{One-loop Feynman diagrams involved in the computation
of the Chern-Simons coefficient $k_{ARR}$. The external line
on top represents a vector $\vecA$ with incoming momentum
$p_0$ and polarisation vector $e_0$. 
The other external lines are gravitons $h$
with incoming momenta $p_1$, $p_2$
and symmetric polarisation tensors
$e_1$, $e_2$.  The internal lines can represent a
massive spin-1/2 fermion, a massive self-dual tensor,
or a massive spin-3/2 fermion.}
\label{diagrams_ARR}
\end{figure}

The presence of quadri-vertices enlarges the family of
one-loop Feynman diagrams that can be built. 
In particular, we have three different topologies,
depicted in figure \ref{diagrams_ARR}.
The total amplitude is given by the sum
\beq \label{summing_diagrams}
2(a) + (b) + 2 (c) \ ,
\eeq
where diagram (a) is counted twice because of the two
possible orientations of the loop,
and diagram (c) is counted twice according to which
graviton is connected to the mixed quadri-vertex.

For each diagram, denominators can be unified
by means of Schwinger parameters. In 
diagram (a) three parameters are needed, as in the previous
section, while diagrams (b) and (c) require only
two parameters. 
Up to minor changes,
the methods described in the previous
section can be applied straightforwardly 
to the diagrams at hand.
In particular, UV divergences in diagrams
(b) and (c) are regulated by means of the replacement
\beq \label{reg_prescription_bis}
 \int_0^\infty d\alpha \int_0^\infty d\beta   \quad
 \rightarrow  \quad
 \int_0^\infty d\alpha \int_0^\infty d\beta   \;
 \theta(\alpha + \beta - \epsilon) \ ,
\eeq
where $\alpha$, $\beta$ are the Schwinger 
parameters and $\epsilon$ is the regulator. 
For the sake of completeness,
we record the two-parameter analog of the identity \eqref{gamma_integral},
\begin{multline}
 \int_0^\infty d\alpha \int_0^\infty d\beta   \; 
 \theta(\alpha + \beta   - \epsilon) 
 \frac{e^{-(\alpha + \beta  )}  }  {(\alpha + \beta  )^a}   \alpha^{n_1} \beta^{n_2}
  = \\
  = \frac{\Gamma(1+n_1) \Gamma(1+n_2)  }{\Gamma(2 + n_1 + n_2  )}
    \Gamma(2 - a + n_1 + n_2  ; \epsilon) \ .
\end{multline}

Let us stress an important difference between the 
present computation and the one discussed in the previous section.
In the case of the gauge Chern-Simons couplings,
the relevant tensor structure \eqref{relevant_AAA}
does not contain any product of external momenta.
This allowed us to use the approximation $\Delta \approx m^2$
in the computation of the diagram in \eqref{dgrm}.
In the present case, one of the two parts of the gauge
invariant tensor structure \eqref{reading_kARR}
is proportional to $p_1 \cdot p_2$. This implies that
we have to keep the $p_1 \cdot p_2$ term
inside $\Delta$ and expand $e^{(\alpha + \beta +\gamma)\ell^2/m^2}$
(or $e^{(\alpha + \beta )\ell^2/m^2}$)
in a power series in the external momenta.
This is indeed crucial to obtain the gauge invariant
structure \eqref{reading_kARR}
after all the three diagrams are combined 
according to \eqref{summing_diagrams}.

As in the case of the gauge Chern-Simons term,
the parity violating part of the diagrams has a better
UV behaviour than expected from naive power-counting.
Nevertheless, the diagrams in which the self-dual tensor and the 
spin-3/2 fermion run in the loop have some divergent parts.
After all diagrams are summed according to 
\eqref{summing_diagrams} and the total expression is
organised in powers of $\epsilon$,
the $\epsilon^0$ coefficient is proportional to the
gauge-invariant combination \eqref{reading_kARR},
while negative-power coefficients are not gauge-invariant.
This leads us to apply a minimal subtraction prescription
and simply drop the unphysical divergent pieces.
In this way the results of table \ref{summary_table}
are obtained.

Let us conclude this section with a side remark.
Recall from section \ref{secAFF} that the relative weight between the
diagram for spin-$1/2$ and spin-$3/2$ fermion contributions
to $k_{AFF}$
 is five. This result can be derived 
straightforwardly from an alternative form of the massive action
for a spin-3/2  $\psi_\mu$,
\beq \label{alternative_action_spin32}
S_{3/2}'
= \int d^5 x \, e  \left[
 -  \bar\psi_\rho \gamma^\mu \cD_\mu \psi^\rho
 + c_{3/2} m \, \bar\psi_\rho  \psi^\rho
  \right] \ , \quad c_{3/2} =\pm 1 \ .
\eeq
Indeed, when this action is evaluated on a flat
background, it gives exactly the same propagator
and vertex as the spin-$1/2$ action \eqref{std_action_spin12},
up to a factor of the metric $\eta_{\mu\nu}$.

Remarkably, the alternative action \eqref{alternative_action_spin32}
gives also the correct relative weight $-19$
between the spin-1/2 and the spin-3/2 contributions to $k_{ARR}$.
This claim has been checked against an explicit 
Feynman diagram computation.
To get the correct result is crucial to take into account
the corrections to the vertices coming from the Christoffel symbols
inside the covariant derivative $\cD_\mu \psi^\rho$.
Indeed, the vertices generated by the Christoffel symbol 
contribute a relative factor of $-24$ that combines with five
times the spin-1/2 result to give $-19$.

This finding resembles a similar result about 
gravitational anomalies in six dimensions \cite{AlvarezGaume:1983ig}.
In order to compute the contribution of a massless chiral spin-$3/2$
field $\psi_\mu$ to gravitational anomalies in six dimensions,
one can use two different Lagrangians, proportional to 
\beq  \label{6d_gravitini}
\bar \psi_\rho \gamma^{\rho\mu\nu} \nabla_\mu \psi_\nu 
\quad \quad \text{or}
\quad \quad
\bar \psi_\rho \gamma^\mu \nabla_\mu \psi^\rho \ ,
\eeq
where $\nabla$ denotes the six-dimensional Levi-Civita 
covariant derivative. It is shown that the 
difference between these Lagrangians cannot affect
the anomalous part of the four-graviton one-loop diagram.
Note that if we compactify the six-dimensional Lagrangians \eqref{6d_gravitini}
on a circle, the resulting actions for the massive Kaluza-Klein modes
have kinetic and mass terms as given in
 \eqref{std_action_spin32} and 
\eqref{alternative_action_spin32}, respectively.
We are thus led to conjecture that 
corrections to the
five-dimensional  Chern-Simons terms \eqref{CS_couplings}
are  insensitive to the precise form
of the differential operator in the kinetic term
and the corresponding form of the mass term.

\subsection{Non-minimal couplings and renormalisation} \label{sec:non_minimal}
 
The aim of this section is to describe the effect of non-minimal couplings
on the Chern-Simons term $\vecA \wedge F \wedge F$,
extending some remarks of \cite{Bonetti:2012fn}.
Gravity is decoupled and the metric is taken to be $\eta_{\mu\nu}$.
As far as fermions are concerned, we consider Pauli couplings
built by contracting a spinor bilinear with the $U(1)$ field strength
$F=dA$. In particular, we have analysed the couplings
\beq \label{non-minimal_fermions}
\cL_{1/2}^{\rm nm} =\tfrac 12 i \tilde q_{1/2} \,F_{\mu\nu}\, \bar\psi \gamma^{\mu\nu} \psi \ , \quad
\cL_{3/2}^{\rm nm} =\tfrac 12 i \tilde q_{3/2} \,F_{\mu\nu}\, \bar\psi_\rho \gamma^{\mu\nu\rho\sigma} \psi_\sigma
+ \tfrac 12 i \tilde q_{3/2}' \, F_{\mu\nu}\, \bar\psi^\mu \psi^\nu \ .
\eeq 
For massive self-dual tensors we have studied instead
\beq \label{non-minimal_tensors}
\cL_{B}^{\rm nm} = \tilde q_B\, \bar B_{\mu\nu} F^{\nu\rho} {B_{\rho}}^\mu 
+ \tilde q_{B}' \, \bar B_{\mu\nu} F^{\nu\rho} B_{\rho\sigma} F^{\sigma \mu} \ .
\eeq

The computation of section \ref{secAFF} can be repeated
including these additional vertices. The corresponding Feynman rules 
can be obtained straightforwardly with the standard prescriptions.
Note, however, that the coupling $\tilde q_B'$ induces a quadri-vertex
and therefore diagrams with a topology as diagram (b) or (c) in figure \ref{diagrams_ARR}
have to be included.

We refrain from a detailed account on the computation.
Nonetheless, its outcome is remarkable: all non-minimal couplings
$\tilde q_{1/2}$ to $\tilde q_B'$ drop from the $\epsilon^0$ coefficient
of the combination of all diagrams and  enter only the coefficients 
of negative powers in $\epsilon$.

This implies that they can be used to cancel divergences 
in the spin-3/2 and tensor diagrams.
Recall from \eqref{finite_and_divergent_32} that the triangle diagram with a 
spin-3/2 fermion running in the loop has two non-vanishing
negative powers of $\epsilon$ if only the minimal coupling $q$ is switched on.
Our computations reveal that turning $\tilde q_{3/2}$, $\tilde q_{3/2}'$ on
does not introduce higher negative powers, i.e.~higher divergences, and does
not affect the coefficient of the $\epsilon^0$ power.
We can thus tune $\tilde q_{3/2}$ and $\tilde q_{3/2}'$
and cancel divergences without altering the finite part of the diagram.

The same strategy can be applied to tensors. The reader might wonder why we take into account
two non-minimal couplings for tensors, if 
the corresponding diagram has only one divergent part, as can be seen from
\eqref{finite_and_divergent_B}. This is necessary since it can be checked that
inclusion of the coupling $\tilde q_B$ introduces higher divergences that require
the introduction of
$\tilde q_{B}'$ to be cancelled.

Our findings suggest the interpretation of non-minimal couplings
\eqref{non-minimal_fermions} and \eqref{non-minimal_tensors} as counterterms.
Dimensional analysis reinforces this claim, since it shows that 
non-minimal couplings $\tilde q_{1/2}$ to $\tilde q_{B}'$ 
have negative mass dimension. In the limit in which the masses of  $\psi$, $B_{\mu\nu}$, and
$\psi_\mu$ tend to infinity and these fields are integrated out,
non-minimal couplings are suppressed.
A similar counterterm analysis for the gravitational Chern-Simons term
is a formidable task and is not addressed in this work. 
Nevertheless, it is plausible that a similar mechanism can be implemented
to cancel all divergences without changing the results
of table \ref{summary_table}.


\section{Consistency with the M-theory to F-theory limit} \label{sec:Mtheory_Ftheory}

In this section we compare the 
field theory results summarised in table \ref{summary_table} 
with the predictions of five-dimensional compactifications of M-theory and F-theory. 
This will enable us to use geometric methods to perform consistency checks of 
the one-loop corrections. Our focus will be 
on the gauge and gravitational Chern-Simons actions in 
five-dimensional low energy effective supergravity theories 
with eight or sixteen supercharges. These arise, 
on the one hand, from M-theory compactifications on 
supersymmetry preserving threefolds. On the other hand, 
they arise from F-theory compactifications to six-dimensions 
preserving $(1,0)$ or $(2,0)$ supersymmetry if the theory is 
further compactified on a circle. The six-dimensional origin 
of five-dimensional effective theories with Chern-Simons terms 
is addressed in \cite{BGH}.

\subsection{Field theory prediction}

Let us apply the results of the one-loop computation
of the previous section to the framework 
of $(1,0)$ and Abelian $(2,0)$ six-dimensional
theories compactified on a circle.
The field content of their supersymmetry multiplets
is summarised in table \ref{spectra_table}
and features chiral fermions and (anti)self-dual
tensors. 
The requirement of anomaly cancellation imposes
some constraints on the 
spectrum of these theories. In particular, the absence 
of gravitational anomalies requires \cite{AlvarezGaume:1983ig, Taylor:2011wt}
\begin{align}
(1,0): & \qquad H-V=273-29\, T \ , \label{anomalies10}  \\
(2,0): & \qquad T=21 \ , \label{anomalies20}
\end{align}
where $T$, $V$, $H$ are 
the numbers 
of tensor multiplets, vector multiplets, and hypermultiplets,
respectively.  

\begin{table}
\centering
\begin{tabular}{| l l   |    l l   |}  
\hline 
\multicolumn{2}{|c|}{\rule[-.3cm]{0cm}{.8cm} $(1,0)$ theory} &   \multicolumn{2}{c|}{$(2,0)$ theory}  \\
 \hline 
\rule[-.3cm]{0cm}{.8cm} gravity multiplet & $(g_{\mu\nu}, \, B^{+}_{\mu\nu} , \, 2\psi^{+}_{\mu} )$    
& gravity multiplet & $(g_{\mu\nu} , \, 5 B^+_{\mu\nu} , \, 4\psi^+_\mu)$  \\
\rule[-.3cm]{0cm}{.8cm} tensor multiplet & $(B^{-}_{\mu\nu} , \,  \phi , \,  2\psi^{-})$    
& tensor multiplet & $(B^-_{\mu\nu} , \, 5\phi , \, 4\psi^-)$  \\
\rule[-.3cm]{0cm}{.8cm}  vector multiplet & $(A_{\mu} , \, 2\psi^{+})$     
&   &    \\
 \rule[-.3cm]{0cm}{.8cm}  hypermultiplet & $(4\phi ,  \, 2\psi^{-})$     
 &   &    \\
\hline
\end{tabular}
 \caption{Schematic form of supersymmetric spectra of $(1,0)$
 and $(2,0)$ theories. 
 The symbols $g_{\mu\nu}$, $B_{\mu\nu}$, $\psi$, $\phi$
 represent the metric, a tensor, a Majorana-Weyl spinor, a real scalar field
 respectively. The prefactor counts the number of fields of a
 given species within each multiplet. 
 The superscript $\pm$ denotes (anti)self-duality
 for the tensors $B$ or chirality for the fermions $\psi$.
   } \label{spectra_table}
\end{table}

Upon compactification on a circle,
the  massive Kaluza-Klein modes of chiral fields
are precisely given by the three families of massive 
fields summarised in table \ref{summary_reps}.
More precisely,
the excited modes of a symplectic Majorana-Weyl spinor
are   Dirac spinors and the modes of a (anti)self-dual
tensor are massive complex self-dual tensors.
We adopt conventions such that  
 a positive chirality in six-dimensions
 correspond to a positive coefficient 
 $c_{1/2}$, $c_B$, or $c_{3/2}$
 in the mass term for excited Kaluza-Klein modes.

The ansatz for the metric reads
\beq
ds^2_6 = ds^2 + r^2 (dy - A^0)^2 \ ,
\eeq
where $r$ is the circle radius and $A^0$
is the Kaluza-Klein vector. This choice of the sign of $A^0$
in the metric ansatz implies that an excited mode with dependence
$e^{iny}$ on the internal coordinate couples minimally to $A^0$
with $U(1)$ covariant derivative $\partial_\mu + i n A^0_\mu$.
This has to be contrasted with the minimal coupling prescription
$\partial_\mu - i q A_\mu$ used in the loop computation.
If we identify $A^0$ and $A$, we infer that $q= -n$
for the $n$-th Kaluza-Klein mode of any six-dimensional field.

In order to compute $k_{AFF}$ and $k_{ARR}$ defined in
\eqref{CS_couplings} we just have to sum
the contributions of table \ref{summary_table}
according to the spectra listed in \ref{spectra_table}.
For a $(1,0)$ theory, we have
\begin{align} \label{summing_spectra_10}
k_{AFF}^{(1,0)} &= - \frac{1}{48 \pi^2} 
\sum_{n = 1}^\infty (-n)^3 \bigg[
2(V - H-T) + 2\cdot 5 + (1-T)(-4)
\bigg] = -\frac{9-T}{24(2\pi)^2} \ ,
   \\
k_{ARR}^{(1,0)} &= - \frac{1}{384 \pi^2} 
\sum_{n = 1}^\infty (-n)  
 \bigg[
2(V - H-T) + 2\cdot (-19) + (1-T)(+8)
\bigg] = \frac{12-T}{24(2\pi)^2} \ ,   \nn
\end{align}
where we made use of the anomaly 
cancellation condition \eqref{anomalies10}
and we
employed  zeta-function regularisations
$\sum n^3 \rightarrow \zeta(-3) = 1/120$ and 
$\sum n \rightarrow \zeta(-1) =- 1/12 $
for the divergent sum over Kaluza-Klein levels.
In a similar fashion, for a $(2,0)$ theory we find
\begin{align} \label{summing_spectra_20}
k_{AFF}^{(2,0)} &= - \frac{1}{48 \pi^2} 
\sum_{n = 1}^\infty (-n)^3 \bigg[
4(-T) + 4\cdot 5 + (5-T)(-4)
\bigg] = 0 \ ,
   \\
k_{ARR}^{(2,0)} &= - \frac{1}{384 \pi^2} 
\sum_{n = 1}^\infty (-n)  
 \bigg[
4( -T) + 4\cdot (-19) + (5-T)(+8)
\bigg] = \frac{T+3}{96(2\pi)^2} = \frac{1}{4(2\pi)^2} \ ,   \nn
\end{align}
where we recalled $T=21$ from \eqref{anomalies20}.
In the next subsection
we will test these field theory predictions
against geometric computation in the framework
of M-theory and F-theory compactifications.

\subsection{F-theory check}

To begin with let us recall some facts about the F-theory
reduction. Considering F-theory on  an elliptically fibered Calabi-Yau threefold $Y_3$ with full
holonomy yields a six-dimensional $(1,0)$ supergravity theory. Non-Abelian 
gauge groups can originate from singularities of the fibre that are also 
singularities of the threefold itself. The singularities of the elliptic fibre
signal the presence of spacetime filling seven-branes. 
In contrast, if one considers F-theory on a two-torus times 
a K3 surface, i.e.~has a trivial elliptic fibration, then no seven-branes 
are present. The six-dimensional effective theory is a $(2,0)$ theory 
with $21$ 
Abelian tensor multiplets.

Since there is no fundamental 
formulation of F-theory in twelve dimensions, the effective action 
of the six-dimensional theories has to be derived via the duality to 
M-theory \cite{Denef:2008wq}. 
This program was carried out for
six-dimensional F-theory vacua in 
\cite{Ferrara:1996wv, Bonetti:2011mw}.
On the one side of the duality,
the low-energy action of M-theory 
is reduced on $Y_3$ or $K3 \times T^2$
down to five dimensions.  
In order for the low-energy action to be a 
viable approximation, the geometry of the 
compactification space has to be smooth.
In particular, possible singularities of $Y_3$ associated
to  non-Abelian gauge groups must be resolved.
At the level of the effective action, this implies that
on the M-theory side of the duality we 
can only probe the five-dimensional Coulomb branch 
of the non-Abelian gauge theory.
On the other side of the duality,
we can use the constraints coming from
$(1,0)$ or $(2,0)$ supersymmetry
to parameterise the sought-for F-theory action
in terms of a few key data.
We then compactify this general six-dimensional action 
on a circle and we compute the low-energy
action for the zero-modes in the Coulomb branch. 
By comparing the five-dimensional theories
on the M-theory and F-theory side of the duality
we can infer the data that determine the 
six-dimensional effective action of F-theory
on $Y_3$ or $K3 \times T^2$.

It is crucial to integrate out all massive fields
in determining the low-energy effective action 
on the F-theory side. This implies that the M-theory
action should encode in particular the information
about the one-loop Chern-Simons couplings $A^0 \wedge F^0\wedge F^0$
and $A^0 \wedge \tr(  R \wedge R)$.
To check this claim we recall some facts about the
reduction of the topological terms of the eleven-dimensional
action on a Calabi-Yau threefold $Y_3$ and on $K3 \times T^2$.

The relevant couplings in the eleven-dimensional action are 
the usual two-derivative Chern-Simons term together with
a topological higher-derivative correction \cite{Vafa:1995fj, Duff:1995wd}.
Suppressing wedge products for brevity, we have
\beq  \label{11d_top_sector}
S^{(11)}_{\rm top} = \int
\bigg[
- \frac {1}{6} \frac{1}{(2\pi)^2} C_3   G_4   G_4
- \frac{1} {192} \frac{1}{(2 \pi)^4} C_3   \left( {\rm tr} R^4 - \frac {1}{4} ({\rm tr} R^2)^2  \right)
\bigg] \ ,
\eeq
where $C_3$ is the M-theory
three-form, with mass dimension 3, $G_4 = dC_3$,  and
$R$ is the curvature two-form.
This form of the action is consistent with the fact that
$\int G_4/(2\pi)$ is  half-integrally quantised and that 
$\exp \, i S $ gives  a well-defined
functional in the path integral, once all 
terms of the effective action and the gravitino
functional measure are taken into account  \cite{Witten:1996md}.
This is crucial to match one-loop computations in field-theory,
since the the standard Feynman rules 
are derived by  an expansion of $\exp \, i S_{\rm int}$,
fixing the absolute normalisation of  one-loop induced Chern-Simons terms.

Let $X_6$ denote the internal space,
for us $Y_3$ or $K3\times T^2$. The M-theory three-form is expanded on a basis $\{\omega_A\}$ of
harmonic two-forms on  $X_6$ as
\beq
C_3 \supset A^A \wedge \omega_A \ ,
\eeq
where $A^A$ are five-dimensional vectors. They have mass dimension
one and  their field strengths $F^A = dA^A$ 
are such that $\int F^A/(2\pi)$
is integrally quantised.
Dimensional reduction of the action \eqref{11d_top_sector}
yields the five-dimensional topological terms \cite{Cadavid:1995bk, Antoniadis:1997eg, Bonetti:2011mw}
\beq
S^{\rm CS} = \frac{1}{(2\pi)^2}
\int \bigg[
- \frac 1 6 \cK_{ABC} A^A F^B F^C
+ \frac {1}{96} c_A A^A {\rm tr} R^2
\bigg] \ ,
\eeq
 where we have introduced
\beq
\cK_{ABC} = \int_{X_6} \omega_A \wedge \omega_B \wedge \omega_C \ , \quad
c_A = \int_{X_6} \omega_A \wedge c_2(X_6) \ . 
\eeq

If $X_6  = Y_3$, the Kaluza-Klein vector
on the F-theory side is matched with the linear
combination of vectors $A^A$ along the direction
of the two-form
\beq
\omega_0 = {\rm PD}(B_2) + \frac 12 c_1(B_2) \ ,
\eeq
where 
 ${\rm PD}(B_2)$ is the Poincar\'e dual two form to the base $B_2$ of the elliptic fibration, and 
$c_1(B)$ is its first Chern class.\footnote{Strictly speaking one has to 
pull back $c_1(B)$ to $Y_3$, but we will suppress the pullback in the following.} 
The geometry of elliptically fibered Calabi-Yau threefolds 
ensures
\begin{align}
\cK_{000} &= \frac 14 \int_{B_2} c_1(B_2)^2 = \frac 14 (10 - h^{1,1}(B_2)) \ , \nn \\
c_0 & = \int_{B_2} \left[ c_2(B_2) + 5c_1(B_2)^2 \right] = 4(13-h^{1,1}(B_2)) \ .
\end{align}
This in turn implies that the Chern-Simons sector
of M-theory on $Y_3$ contains the terms
\beq
S^{\rm CS} \supset \frac{1}{(2\pi)^2}
\int \bigg[
- \frac {10 - h^{1,1}(B_2)}{24} A^0 F^0 F^0
+ \frac {13 - h^{1,1}(B_2)}{24}  A^0 {\rm tr} R^2
\bigg] \ .
\eeq
We just have to recall that the number of tensor multiplets
of the $(1,0)$ theory is related to the geometry of $Y_3$ by
\beq
h^{1,1}(B_2) = T+1
\eeq
to recognise a perfect match with the 
field theory prediction of the previous subsection.

In the case of compactification of M-theory
on $X_6 = K3 \times T^2$, the Kaluza-Klein
vector is identified with the vector 
along the only two-form on the torus, which we denote $\omega_0$.
As a result,
\beq
\cK_{000} = 0 \ , 
\quad
c_0 = \int_{K3\times T^2} \omega_0 \times c_2(K3 \times T^2) = \int_{K3} c_2(K3) = 24 \ .
\eeq
This implies that the gauge Chern-Simons term is absent,
while the gravitational Chern-Simons 
is given by
\beq
S^{\rm CS} \supset \frac{1}{(2\pi)^2}
\int 
 \frac {1}{4}  A^0 {\rm tr} R^2
\ ,
\eeq
in agreement with
the field theory computation.

So far we have focused on Chern-Simons coupling involving 
only the Kaluza-Klein vectors. There are additional
terms in the reduction of M-theory on $Y_3$ that
are interpreted as one-loop effects on the F-theory side.
They are of the form
\beq
k_{0ij} \int A^0 F^i F^j + k_{ijk} \int A^i F^j F^k \ ,
\eeq
where $A^i$ are the five-dimensional vectors that are lifted to 
six-dimensional vectors. The index $i$ labels the Cartan generators
of the gauge group, since the duality between M-theory and
F-theory only works in the Coulomb phase.
The coefficients $k_{0ij}$, $k_{ijk}$
can be computed geometrically
and are related to the charged spectrum of the theory,
see for instance \cite{Intriligator:1997pq, Grimm:2011fx}.

To compute the coefficient of these couplings in field theory
we need to consider diagrams where all massive fields charged 
under $A^0$ and/or $A^i$ run. Those are the Kaluza-Klein 
zeromodes and excited modes of 
the fields that acquire a mass after the gauge group
is broken by giving a non-vanishing VEV to
the scalars in the five-dimensional vector multiplets.
We do not perform here a similar analysis,
but we are confident about the applicability of the techniques
developed so far to attack this problem.
Note that it has indeed been shown in \cite{Intriligator:1997pq}
that the Chern-Simons coefficient $k_{ijk}$
receives one-loop corrections by
massive gauge degrees of freedom.

Let us close this section with a comment about a special case that
recently attracted interest \cite{KashaniPoor:2013en}.
Namely, let us consider an M-theory compactification with $\chi(Y_3)
=0$. When $Y_3$ is elliptically
fibered one can lift the theory to 
a six-dimensional $(1,0)$ model.
For simplicity, we assume that $Y_3$ has no gauge group singularities
and hence the $(1,0)$ theory has no 
vector multiplets,
$V=0$. In this case the Euler number is
simply given by $\chi = -60 \int_{B_2} c_1(B)^2 = -60 (9-T)$ and we
see that  
$\chi=0$ implies $T=9$. The anomaly
cancellation condition \eqref{anomalies10} 
requires then 
 $H=12$.
Can this model be interpreted as a 
spontaneously broken 
$(2,0)$ theory?
Suppose we are given a possibly non-Abelian $(2,0)$ theory
with  21 tensor multiplets, in accord with
absence of gravitational anomalies.
They   correspond
to 21 tensor multiplets and 21 hypermultiplets 
in $(1,0)$ language,
as can be seen from table \ref{spectra_table}.
Let us further imagine that the original
theory undergoes a spontaneous supersymmetry breaking
in such a way that only   $T$ tensor multiplets out of 21 and
and only $H$ hypermultiplets out of 21  remain massless.
In order for the resulting $(1,0)$ theory 
to be free of gravitational anomalies,
we must have $H= 273 - 29 T$. 
The requirement $0 \le H \le 21$ together
with the integrality of $T$ determines $T=9$,
$H=12$ as the only possible breaking pattern.
This agrees with the geometric setup with $\chi = 0$.
Furthermore,
for $T=9$ we have $k_{AFF}^{(1,0)} = 0$, see \eqref{summing_spectra_10},
and the term $A^0 \wedge F^0 \wedge F^0$,
which is incompatible with 16 supersymmetries,
does not enter the circle reduction of the $(1,0)$ theory.
These might be considered as
hints in favour of the spontaneous symmetry breaking 
scenario. If  such breaking
is actually possible, and how it may be realised, remains to be investigated.


   \section{Dual heterotic string on $K3 \times S^1$}\label{sec:heterotic} 

In this section we compare the gauge theory loop corrections 
derived in section \ref{field_theory_section} with a string loop computation 
performed in the heterotic string theory on $K3 \times S^1$. 
The outcome of such computation
only depends on topological data of  $K3$.
A match with the F-theory setup of the previous section
is expected on the basis of heterotic/F-theory
duality for elliptically fibered $K3$.
 We recall the basic five-dimensional setup in section \ref{Sect:HetSetup} and present 
explicit one-loop corrections to the action in section \ref{het_oneloop_1}.

\subsection{Heterotic setup}\label{Sect:HetSetup}
The spectrum of heterotic string theory compactified on $K3$ consists of a 
single tensor multiplet, $V$ vector multiplets and $H$ hypermultiplets 
coupled to the six-dimensional supergravity multiplet. The number of tensor multiplets 
is larger if we consider $K3$ manifolds with singularities, however, we will not 
discuss this possibility.

Compactifying this theory further on $S^1$ (see \cite{Bonetti:2011mw}), the five-dimensional 
effective action contains $V^{(5)}=V+2$ physical vector multiplets and $H^{(5)}$ hypermultiplets 
coupled to the supergravity multiplet.  The additional two vector multiplets 
come from the reduction of the single tensor multiplet and the six-dimensional 
supergravity multiplet. The scalar fields in these multiplets are the five-dimensional 
dilaton $\phi$ and the radius of $S^1$ which we will denote by $\rad$. The vector 
fields in turn correspond to the KK vectors coming from the reduction on $S^1$, i.e. 
from the reduction of the metric ($g_{\mu 6}$) and the anti-symmetric B-field ($b_{\mu6}$) 
respectively. We follow \cite{Antoniadis:1995vz} and denote these two as
\beq \label{ABbasis}
A_\mu=g_{\mu6}+b_{\mu6}\ , \qquad
 B_\mu=g_{\mu6}-b_{\mu6} \ ,
\eeq
with their field strength denoted $F_{A,B}$ respectively. Finally, we denote the curvature 2-form $
R=d\omega+\omega\wedge \omega$, where $\omega$ is the spin-connection. $R$ is related to the Riemann tensor in the usual manner
\begin{align}
{R^a}_b=\frac{1}{2}\,{\mathcal{R}^{a}}_{bcd}\mathfrak{e}^c\wedge\mathfrak{e}^d\,,&&\text{with the vielbein}&&\mathfrak{e}^a=e^a_\mu dx^\mu\,,
\end{align}
which will turn out to be more convenient for explicit calculations later on.
\subsection{String amplitudes}\label{het_oneloop_1}
\subsubsection{Vertex operators}
Within the above setting we will particularly be interested to the one-loop corrections to the following terms in the string effective action
\begin{align}
\mathcal{L}_{\text{ST}}=
&+\tfrac{1}{2}\,\mathcal{F}^{(A\mathcal{R}\mathcal{R})}(\rad)\,\epsilon^{\lambda\mu\nu\rho\tau}\,A_\lambda {\mathcal{R}^{\alpha}}_{\beta\mu\nu}\,{\mathcal{R}^\beta}_{\alpha\rho\tau}+
\tfrac{1}{2}\, \mathcal{F}^{(B\mathcal{R}\mathcal{R})}(\rad)\,\epsilon^{\lambda\mu\nu\rho\tau}\,B_\lambda {\mathcal{R}^{\alpha}}_{\beta\mu\nu}\,{\mathcal{R}^\beta}_{\alpha\rho\tau}\nonumber\\
&+\tfrac{1}{6}\mathcal{F}^{(AF_AF_A)}(\rad)\,\epsilon^{\lambda\mu\nu\rho\tau}A_\lambda\, F_{A,\mu\nu}\,F_{A,\rho\tau}+\tfrac{1}{2}
\mathcal{F}^{(BF_AF_A)}(\rad)\,\epsilon^{\lambda\mu\nu\rho\tau}B_\lambda\, F_{A,\mu\nu}\,F_{A,\rho\tau}\,,\label{EffectiveStringCouplings}
\end{align}
where all coupling functions $\mathcal{F}$ are functions of the radius $\rad$. We will compute the corresponding one-loop scattering amplitudes using the RNS-formalism. In this approach the former are integrals over (higher-genus) Riemann surfaces and the external states are represented through emission vertex operators inserted at punctures, whose position needs to be integrated over the entire world-sheet. Thus, the first 
step is to discuss the precise form of the vertex operators for all states involved. We will 
use the same notation as in \cite{Antoniadis:1995vz} and denote by $X^\mu(z)$ the embedding 
coordinates for the five spacetime directions and $X^6(z)$ the coordinate of the circle, 
while $(\psi^\mu(z),\psi^6(z))$ are the corresponding (left-moving) superpartners. Here $z$ 
are two-dimensional coordinates on the world-sheet. With this notation, the graviton vertex operator in the $0$ ghost picture is
\begin{align}
V^{(0)}_{\mathcal{R}}(h,p;z)=h_{\mu\nu}\left[\partial X^\mu+i(p\cdot \psi) \psi^\mu\right]\,\bar{\partial} X^\nu\,e^{ip\cdot X}\,,
\end{align}
which is characterised by a symmetric, traceless polarisation tensor $h_{\mu\nu}$ and a 
five-momentum $p_\mu$ such that $p^\mu h_{\mu\nu}=0$. 
The KK vector fields have vertex operators
\beq
V^{(-1)}_A(\epsilon;z)=\epsilon_\mu e^{-\varphi}\psi^\mu\bar{\partial} X^6\,e^{ip\cdot X}\,,
\qquad
 V^{(-1)}_B(\epsilon;z)=\epsilon_\mu e^{-\varphi}\psi^6\bar{\partial} X^\mu\,e^{ip\cdot X} \,,
\eeq
and are determined by the polarisation $\epsilon_\mu$. These vertices are written in the $(-1)$ 
picture, referring to the ghost system on the string world-sheet, which we have bosonised in 
terms of the scalar $\varphi$. In order to balance the ghost charge in a given amplitude, we also need picture changing 
operators (PCO), the relevant part of which is given by
\beq
V_{\text{PCO}}=e^{\varphi} T_F\,, 
\qquad
T_F=\psi_\mu\partial X^\mu+\psi^6\partial X^6+T_{F}^{\text{int}}\,.
\eeq
With these expressions we are in a position to compute explicit amplitudes.
\subsubsection{World-sheet CFT}\label{Sect:LoopComputations}
The effective couplings (\ref{EffectiveStringCouplings}) can be related to explicit string amplitudes in the following manner
{\allowdisplaybreaks
\begin{align}
&\mathcal{F}^{(A\mathcal{R}\mathcal{R},B\mathcal{R}\mathcal{R})}(\rad)\nonumber\\
&=\int \frac{d^2\tau}{\tau_2^2}\int d^2z_{1,2,3}\left\langle V^{(0)}_{\mathcal{R}}(h^{(1)},p^{(1)};z_1) V^{(0)}_{\mathcal{R}}(h^{(2)},p^{(2)};z_2) V^{(-1)}_{A,B}(\epsilon,p;z_3)\,V_{\text{PCO}}(r_0)\right\rangle\,,\nonumber\\
&\mathcal{F}^{(AF_AF_A,BF_AF_A)}(\rad)\nonumber\\
&=\int \frac{d^2\tau}{\tau_2^2}\int d^2z_{1,2,3}\left\langle V^{(0)}_{A}(\epsilon^{(1)},p^{(1)};z_1) V^{(0)}_{A}(\epsilon^{(2)},p^{(2)};z_2) V^{(-1)}_{A,B}(\epsilon^{(3)},p^{(3)};z_3)\,V_{\text{PCO}}(r_0)\right\rangle,\label{AmpsVert}
\end{align}
}
where the integral over $\tau=\tau_1+i\tau_2$ runs over the fundamental domain of 
the world-sheet torus. Notice also that the position of the PCO ($r_0$) is not 
integrated over, since the full correlator is independent of $r_0$. Since we work 
at one-loop, in principle, we have to take into account all different spin-structure 
configurations, corresponding to all possible boundary conditions of the world-sheet 
fermions along the two cycles of the torus. Fortunately, for these Chern-Simons like couplings only the odd spin-structure will be of relevance, since they involve 
contractions with $\epsilon^{\mu\nu\rho\sigma\tau}$. 
The correlators in (\ref{AmpsVert}) are understood within the full world-sheet theory. 
To calculate them, we first have to discuss the structure of the internal CFT where we 
essentially follow \cite{Antoniadis:2009nv} (see also \cite{Hohenegger:2011us}). The 
partition function of $S^1$ is captured by the theta-series of the $\Gamma^{(1,1)}$ Narain lattice, whose momenta 
will be denoted $(P_L,P_R)$ and which are functions of the radius $\rad$ only. Concerning the 
remainder of the internal theory, the amplitudes (\ref{AmpsVert}) fortunately are not sensitive 
to the full details of the $K3$ compactification, which would make it prohibitively difficult 
to compute them in general. In fact the only contribution is independent of the spin-structure and in the notation of \cite{Antoniadis:2009nv} can be written in the form
\begin{align}
\bar{F}(\bar{\tau}):=d_0\,\frac{\bar{E}_4(\bar{\tau})\bar{E}_6(\bar{\tau})}{\bar{\eta}^{24}}\,,
\end{align}
where we have also included the partition functions of the internal fermions and bosons.
Here $\bar{E}_{2k}$ for $k\geq2$ are the anti-holomorphic 
Eisenstein series of weight $2k$. The normalisation constant $d_0$ (which is not fixed through K\"ocher's principle alone) 
will be left undetermined in our computation and could be determined later by comparison to vector field amplitudes.
\subsubsection{Explicit amplitudes}
We now have all the ingredients to compute the amplitudes (\ref{AmpsVert}). By performing all contractions of the world-sheet fields, we find 
\begin{align}
&\mathcal{F}^{(A\mathcal{R}\mathcal{R})}=\frac{\pi^3}{3}\epsilon_\lambda\, p_{\alpha}^{(1)}p_{\beta}^{(2)}p^{(1)}_\tau p^{(2),[\nu}h^{(1)}_{\mu\nu}h^{(2),\tau]}_\rho\epsilon^{\alpha\mu\beta\rho\lambda}\,\rad\partial_\rad\,\mathcal{I}(\rad)\,,\nonumber\\
&\mathcal{F}^{(B\mathcal{R}\mathcal{R})}=\frac{\pi^3}{3}\epsilon_\lambda\, p_{\alpha}^{(1)}p_{\beta}^{(2)}p^{(1)}_\tau p^{(2),[\nu}h^{(1)}_{\mu\nu}h^{(2),\tau]}_\rho\epsilon^{\alpha\mu\beta\rho\lambda}\,\mathcal{I}(\rad)\,,\nonumber\\
&\mathcal{F}^{(AF_AF_A)}=64\pi^4d_0\, \epsilon^{(1)}_\mu\epsilon^{(2)}_\nu\epsilon^{(3)}_\rho p^{(1)}_\alpha p^{(2)}_\beta \epsilon^{\mu\alpha\nu\beta\rho}\,\rad\partial_\rad\mathcal{I}_0(\rad)\,,\nonumber\\
&\mathcal{F}^{(BF_AF_A)}=64\pi^4d_0\, \epsilon^{(1)}_\mu\epsilon^{(2)}_\nu\epsilon^{(3)}_\rho p^{(1)}_\alpha p^{(2)}_\beta \epsilon^{\mu\alpha\nu\beta\rho}\,\mathcal{I}_0(\rad)\,,
\end{align}

Here we have introduced the integral over the world-sheet torus
\begin{align}
&\mathcal{I}_0(\rad):=\frac{i}{16\pi^2 d_0}\int \frac{d^2\tau}{\tau_2^2}\,F(\bar{\tau})\partial_{\bar{\tau}}\left[\tau_2^{1/2}\sum_{\Gamma^{(1,1)}}q^{\frac{1}{2}P_L^2}\bar{q}^{\frac{1}{2}P_R^2}\right]=\frac{\theta(1-\rad)\rad^3}{3}+\frac{\theta(\rad-1)}{3\rad^3}\,,\label{AFTintegral}\\
&\mathcal{I}(\rad):=\int \frac{d^2\tau}{\tau_2^{3/2}}\,\hat{\bar{E}}_2(\bar{\tau})\bar{F}(\bar{\tau})\sum_{\Gamma^{(1,1)}}q^{\frac{1}{2}P_L^2}\bar{q}^{\frac{1}{2}P_R^2}=\nonumber\\
&\hspace{0.2cm}-8d_0\pi\,\left[6\left(\rad+\frac{1}{\rad}\right)+5\left(\rad\theta(1-\rad)+\frac{\theta(\rad-1)}{\rad}\right)-2\left(\rad^3\theta(1-\rad)+\frac{\theta(\rad-1)}{\rad^3}\right)\right]\,,\label{LatticeIntegral}
\end{align}
where $\hat{E}_2(\tau,\bar{\tau})=E_2(\tau)-3/(\pi\tau_2)$ is the quasi-holomorphic 
second Eisenstein series. $\mathcal{I}_0(r)$ was already evaluated in \cite{Antoniadis:1995vz}, while $\mathcal{I}(r)$ is performed in appendix~\ref{App:Torus}.


\subsection{Change of basis}
In order to make contact to \cite{Antoniadis:1995vz} we will now make a change 
of basis of the vector fields. So far, in order to keep the computation of the 
loop amplitudes as simple as possible we have considered the vector fields in 
the basis (\ref{ABbasis}). In order to physically interpret the couplings, however, 
we will now return to the basis $(g_{\mu6},b_{\mu6})$. The first couplings to be 
rewritten are the Chern-Simons terms 
\begin{align}
\mathcal{L}^{\text{1-loop}}_{AFF}&=\frac{1}{3!}\,\mathcal{F}^{(AF_AF_A)}(A_\mu dx^\mu)\wedge F_A\wedge F_A+\frac{1}{2!}\,\mathcal{F}^{(BF_AF_A)}(B_\mu dx^\mu)\wedge F_A\wedge F_A\nonumber\\
&=\frac{2\pi^4 d_0}{3}\left[\rad\partial_\rad\mathcal{I}_0(r)(A_\mu dx^\mu)+3\mathcal{I}_0(r) (B_\mu dx^\mu)\right]\wedge F_A\wedge F_A\,.\label{AFFig}
\end{align}
Upon changing to the basis
\beq  \label{BasisChange}
A_\mu=A^{(1)}_\mu/\rad-A^{(2)}_\mu \rad\,, 
\qquad
B_\mu=A^{(1)}_\mu/\rad+A^{(2)}_\mu \rad\,,
\eeq
the coupling (\ref{AFFig}) becomes\footnote{Notice that the explicit dependence on the compactification radius $r$ is $S^1$ is due to the fact that we are working in a frame in which the anti-symmetric two-form field has not been dualised into a vector.}
\begin{align}
\mathcal{L}^{\text{1-loop}}_{AFF}&=64\pi^4d_0\left[a_1(A^{(1)}_\mu dx^\mu)+2\left(-a_1\rad^2+\frac{a_2}{2\rad^4}\right)(A^{(2)}_\mu dx^\mu)\right]\wedge F^{(1)}\wedge F^{(1)}+\nonumber\\
&+64\pi^4d_0\left[a_2(A^{(2)}_\mu dx^\mu)+2\left(-\frac{a_2}{\rad^2}+\frac{a_1\rad^4}{2}\right)(A^{(1)}_\mu dx^\mu)\right]\wedge F^{(2)}\wedge F^{(2)}\,,
\end{align}
where we have introduced (see \cite{Antoniadis:1995vz})
\beq
 a_1=\theta(1-\rad)/3\,,
 \qquad
  a_2=\theta(\rad-1)/3\,.
\eeq
Similarly, we can also consider
\begin{align}
\mathcal{L}^{\text{1-loop}}_{F\mathcal{R}\mathcal{R}}&=\frac{1}{2!}\,\mathcal{F}^{(A\mathcal{R}\mathcal{R})}(A_\mu dx^\mu)\wedge\text{Tr}(\mathcal{R}\wedge\mathcal{R})+\frac{1}{2!}\,\mathcal{F}^{(B\mathcal{R}\mathcal{R})}(B_\mu dx^\mu)\wedge\text{Tr}(\mathcal{R}\wedge\mathcal{R})\nonumber\\
&=-\frac{\pi^3}{3}\left[\rad\partial_\rad\mathcal{I}(\rad)\,(A_\mu dx^\mu)+\mathcal{I}(\rad)\,(B_\mu dx^\mu)\right]\wedge\text{Tr}(\mathcal{R}\wedge\mathcal{R})\,.\label{GravChernSim}
\end{align}
After the change of basis (\ref{BasisChange}) this becomes
\begin{align}
\mathcal{L}^{\text{1-loop}}_{F\mathcal{R}\mathcal{R}}&=-\frac{8\pi^4d_0}{3}\left[6+15a_1+6\left(-2\rad^2a_1+\frac{a_2}{\rad^4}\right)\right](A^{(1)}_\mu dx^\mu)\wedge\text{Tr}(\mathcal{R}\wedge\mathcal{R})\nonumber\\
&-\frac{8\pi^4d_0}{3}\left[6+15a_2+6\left(-\frac{2a_2}{\rad^2}+a_1\rad^4\right)\right](A^{(2)}_\mu dx^\mu)\wedge\text{Tr}(\mathcal{R}\wedge\mathcal{R})\,.
\end{align}
Extracting an overall factor of $-256\pi^6d_0$, which is common to both couplings\footnote{As already mentioned, we are not interested in the total overall normalisation of the effective action.}, we can read off
\beq
 k_0=-\frac{1}{12\pi^2}\,,
 \qquad
 \kappa_0=\frac{11}{96\pi^2}\,,
\eeq
in agreement with (\ref{summing_spectra_10}) for $T=1$, suitable for the heterotic compactification under study.


 \section{Conclusions}
In this paper we have studied Chern-Simons terms in five-dimensional gauge theories with non-Abelian gauge groups. More precisely we have considered terms of the form $A^0\wedge F^0\wedge F^0$ and $A^0\wedge \text{tr}(R\wedge R)$ where $A^0$ is a massless $U(1)$ gauge field with field strength $F^0$ and $R$ is the curvature two-form. These terms are interesting for a number of reasons. Indeed, they appear as one-loop corrections to the low energy effective action by integrating out massive excited modes beyond a certain cutoff scale. While such contributions are usually suppressed in the limit of large cutoff and are therefore generically neglected, these Chern-Simons terms are a rare example of a class of couplings that are independent of the scale introduced by the cutoff and thus have to be included. This property makes them very interesting for a number of applications: Foremost, in a setting where the five-dimensional theory is obtained through compactification of a six-dimensional theory on a circle (and $A^0$ is identified with the Kaluza-Klein vector), we observe that these couplings encode information about the higher-dimensional gravitational anomalies. Thus, the Chern-Simons terms can be used as a tool to test whether a possible higher-dimensional parent theory is plagued by anomalies. We have more to say about the exciting possibility to use this connection to probe the quantum-consistency of a variety of five-dimensional gauge theories in an upcoming paper~\cite{BGH}.

To compute these terms we have used three different approaches: First we have directly calculated them at the one-loop level in field theory. In this approach we have worked out the Feynman diagrams for massive spin-1/2, massive spin-3/2, and massive two-form tensors running in the loop. While these amplitudes are generically divergent, we have argued that it is possible to consistently introduce counterterms  to get rid of all divergences. In the case of the gauge Chern-Simons term $A^0\wedge F^0\wedge F^0$, we have even made this procedure explicit. In a second step, we have successfully compared the field theoretic results with the predictions of five-dimensional M- and F-theory compactifications giving rise to theories with eight or sixteen supercharges. In this case, the couplings can be obtained from purely geometric considerations determining the spectrum of the five-dimensional effective theory. In a final step, using the duality between F-theory and heterotic string theory, we have re-obtained the couplings through a one-loop computation in heterotic string theory compactified on $K3\times S^1$, where $K3$ is realised as an elliptic fibration. Assuming that the latter has no singularities, the spectrum is limited to include only a single tensor multiplet. While we believe that it should in principle be possible to include more tensor multiplets, we have not further investigated this question in the current work. All three approaches in the end yield the same answer, which lends strong support to our understanding of Chern-Simons terms at the quantum level.

We expect further applications of our results in other fields of physics. For example, recent progress towards a description of the world-volume theory of multiple coincident M5-branes has been achieved by studying five-dimensional non-Abelian gauge theories including massive Kaluza-Klein modes \cite{Bonetti:2012st,Ho:2011ni,Huang:2012tu}. We expect that Chern-Simons terms play an important role in determining the quantum consistency of these approaches and may also be relevant in extracting data of the M5-brane theory that are robust under dimensional reduction. For example, the conformal anomaly of $N$ coincident M5-branes shows a typical scaling behaviour of the form $N^3$ as was established using methods of the dual supergravity theory \cite{Bastianelli:2000hi}, or more recently from matrix model approaches 
\cite{Kallen:2012va}
 or other field theoretic methods \cite{Yi:2001bz,Maxfield:2012aw}.
 Non-perturbative topological string theory is yet another
 approach to  study  coincident M5-branes \cite{Lockhart:2012vp}. We expect that this scaling behaviour can also be extracted by studying certain classes of Chern-Simons terms involving a background metric (see \cite{Bonetti:2012st} for some preliminary discussion).

Another possible application concerns higher-derivative corrections to black hole entropy in supergravity and string theory. Indeed, as discussed in \cite{Castro:2007sd,Castro:2007hc,Castro:2007ci,Castro:2008ne}, a gauge-gravitational Chern-Simons term of the form $A \wedge \tr(R \wedge R)$ (or rather its supersymmetric completion along the lines of \cite{Hanaki:2006pj}) plays a crucial role for the entropy and the attractor behaviour of five-dimensional black objects. Upon further compactification on a circle, these are also relevant four dimensional black holes. We thus expect our results to also be relevant for future studies of quantum corrections to black holes, in the supersymmetric as well as non-supersymmetric case.

Finally, we believe that results are not limited to five dimensions alone. Indeed, we expect similar relations between Chern-Simons terms in odd-dimensional theories to higher-dimensional anomalies. For some recent examples in three dimensions see e.g.~\cite{Cvetic:2012xn}. In the spirit of \cite{BGH}, Chern-Simons terms might be a useful tool in classifying a subclass of quantum consistent theories in any number of dimensions.

\section*{Acknowledgements}
We would like to thank Luis Alvarez-Gaum\'{e}, Ignatios Antoniadis, Sophia Borowka, Thomas Hahn, Henrik Johansson, Denis Klevers, Neil Lambert, Noppadol Mekareeya, Boris Pioline, Tom Pugh, Raffaele Savelli, and Washington Taylor for useful discussions. The work of FB and TG was supported by a grant of the Max-Planck society. SH is grateful to the Ludwig-Maximilians-University and the Max-Planck Institute for Physics in Munich for kind hospitality during completion of this work.


 \appendix
 
  \section{Notations and conventions} \label{app_notations}
 
We use Greek letters $\mu,\nu,\dots$ for curved spacetime indices and
Latin letters $a,b,\dots$ for flat spacetime indices. 
They both take values $0,\dots,4$.
(Anti)symmetrisation on any kind of indices 
is performed with  weight one, e.g.~$X_{(\mu\nu)} = \tfrac 12 X_{\mu\nu} + \tfrac 12 X_{\nu\mu}$.
The metric $g_{\mu\nu}$, the vielbein ${e^a}_{\mu}$,
and the flat metric $\eta_{ab}$ are such that
\beq
g_{\mu\nu} = \eta_{ab} {e^a}_\mu {e^b}_\nu \ , \quad
\eta_{ab} = {\rm diag} (-,+,+,+,+) \ .
\eeq
We adopt the
following conventions for the Christoffel symbol and the curvature tensors,
\begin{align} \label{Christoffel_and_Riemann}
{\Gamma^\rho}_{\mu\nu} & = 
\tfrac 12 g^{\rho\sigma} \left( 
\partial_{\mu} g_{\nu\sigma} + \partial_\nu g_{\mu \sigma} - \partial_\sigma g_{\mu\nu}
\right)  \ , \nn \\
{R^\lambda}_{\tau\mu\nu} & = 
\partial_\mu {\Gamma^\lambda}_{\nu\tau} - \partial_\nu {\Gamma^\lambda}_{\mu\tau}
+ {\Gamma^\lambda}_{\mu\sigma} {\Gamma^\sigma}_{\nu\tau}
- {\Gamma^\lambda}_{\nu\sigma} {\Gamma^\sigma}_{\mu\tau} \ .
\end{align}

The Levi-Civita tensor is denoted $\epsilon_{\mu\nu\rho\sigma\tau}$
and is such that
\beq
\epsilon_{01234} = \sqrt{-g}  \quad 
\text{for curved indices  $0,\dots,4$}\ ,
\eeq
where $g = \det{g_{\mu\nu}}$. 
The wedge product, Hodge star, and exterior derivative of
differential
forms satisfy
\begin{gather}
(\alpha \wedge \beta)_{\mu_1 \dots \mu_{p+q}} = 
\tfrac{(p+q)!}{p!q!} \alpha_{[\mu_1 \dots \mu_p} \beta_{\mu_{p+1} \dots \mu_{p+q}]} \ ,
\quad
(d\alpha)_{\mu_0 \dots \mu_p} = (p+1) \partial_{[\mu_0} \alpha_{\mu_1 \dots \mu_p]} \ , \nn \\
(*\alpha)_{\mu_1 \dots \mu_{5-p}} = 
\tfrac{1}{p!} \, \alpha^{\nu_1 \dots \nu_p} \epsilon_{\nu_1 \dots \nu_p \mu_1 \dots \mu_{5-p}} \ ,
\end{gather}
in which $\alpha$ is $p$-form and $\beta$ is a $q$-form.

The spin connection $\omega_{\mu ab} = \omega_{\mu[ab]}$ is determined
by the vielbein ${e^a}_\mu$ according to the torsionless condition 
\beq
de^a + {\omega^a}_b \wedge e^b = 0 \ .
\eeq
The curvature two-form is given by
\beq
{R^a}_b = d{\omega^a}_b + {\omega^a}_c \wedge {\omega^c}_b = \tfrac 12 {e^a}_\lambda {e_b}^\tau {R^\lambda}_{\tau\mu\nu} \, dx^\mu \wedge dx^\nu \ ,
\eeq
where in the last expression ${R^\lambda}_{\tau\mu\nu}$ denotes the Riemann tensor 
defined in \eqref{Christoffel_and_Riemann}.

Five-dimensional gamma matrices $\gamma^a$ are constant, complex-valued $4 \times 4$
matrices satisfying the anticommutation relation
\beq
\{ \gamma^a , \gamma^b \} = \eta^{ab} \ .
\eeq
We always make use of the shorthand notation
$\gamma^{a_1 \dots a_p} = \gamma^{[\mu_1} \dots \gamma^{\mu_p]}$
and we work in a representation of gamma matrices such that
\beq
\gamma^{abcde} = i\, \epsilon^{abcde} \, \mathbb{I}_4 \ .
\eeq
The matrix $\gamma^0$ is taken to be anti-Hermitian,
while $\gamma^1, \dots\, \gamma^4$ are Hermitian. The Dirac
conjugate of a spinor $\psi$ is defined as
\beq
\bar\psi = \psi^\dagger \gamma^0 \ .
\eeq
Our spinors are Grassmann variables. 
Note that 
in our convention complex conjugation
acts on products of Grassmann variables according to $(ab)^* = a^* b^*$.

 \section{Gravitational perturbative expansion}  \label{app_pert_gravity}

In this appendix we record some useful identities about the
gravitational perturbative expansion   
around flat spacetime.
More precisely, we assume a metric of the form
\beq
g_{\mu\nu} = \eta_{\mu\nu} + h_{\mu\nu} \ ,
\eeq
and compute some geometrical quantities derived from the metric
in a formal power series in $h_{\mu\nu}$. 
On the right hand side of the following identities,
indices are raised and lowered with the flat metric $\eta_{\mu\nu}$
and its inverse. For instance, $h^{\mu\nu} = h_{\lambda \tau} \eta^{\lambda \mu} \eta^{\tau \nu}$.

The total inverse metric  
and volume form are given by
\begin{align}
g^{\mu\nu} & = 
\eta^{\mu\nu} - h^{\mu\nu} + h^{\mu\lambda} {h_{\lambda}}^{\nu} + \cO(h^3) \ , \nn \\
\sqrt{-g} & = 1 + \tfrac 12 {h^\mu}_\mu + \tfrac 18 ({h^\mu}_\mu {h^\nu}_\nu - 2 h^{\mu\nu} h_{\mu\nu}) + \cO(h^3) \ .
\end{align}
The Christoffel symbols and the Riemann tensor are expanded as
\begin{align}
{\Gamma^\rho}_{\mu\nu} & = \tfrac 12 (\eta^{\rho\sigma} - h^{\rho \sigma})
(\partial_{\mu} h_{\nu\sigma} + \partial_{\nu} h_{\mu\sigma} - \partial_{\sigma} h_{\mu\nu})
+ \cO(h^3) \ , \nn \\
g_{\rho\tau} {R^{\tau}}_{\sigma \mu\nu} & =\Big[ - \tfrac 12 \partial_{\rho} \partial_{\mu} h_{\sigma \nu}
 - \tfrac 18 \partial_\lambda h_{\mu\rho} \partial^\lambda h_{\nu\sigma}
+ \tfrac 18 \partial_\mu h_{\sigma \lambda} \partial_\nu {h_{\rho}}^\lambda
- \tfrac 14 \partial_\rho h_{\mu \lambda} \partial_\nu {h_{\sigma}}^\lambda 
  + \tfrac 18 \partial_\rho h_{\nu \lambda} \partial_\sigma {h_{\mu}}^\lambda \nn \\
& + \tfrac 14 \partial_\mu h_{\rho \lambda} \partial^\lambda h_{\nu \sigma}
- \tfrac 14 \partial_\rho h_{\nu \lambda} \partial^\lambda h_{\mu \sigma} 
- (\mu \leftrightarrow \nu) \Big] - (\rho \leftrightarrow \sigma) + \cO (h^3) \ .
\end{align} 
 
In order to couple spinors to gravity we need to introduce a vielbein ${e^a}_\mu$.
It is determined by the metric only up to local Lorentz transformations.
We fix this ambiguity by imposing the gauge condition
\beq
\eta_{ab} {\delta^b}_{[\lambda} {e^a}_{\mu]} = 0 \ ,
\eeq
where the Kronecker delta plays the role
of the vielbein for the flat metric $\eta_{\mu\nu}$.
We thus find the following expansions of the vielbein and its inverse,
\begin{align}
{e^a}_{\mu} & = \eta^{ab} {\delta_b}^\lambda \big[ \eta_{\lambda \mu}
+ \tfrac 12 h_{\lambda \mu} - \tfrac 18 h_{\lambda \tau} {h_{\mu}}^\tau
+ \cO(h^3)
\big] \ , \nn \\
{e_a}^\mu & = \eta_{ab} {\delta^b}_\lambda \big[
\eta^{\lambda \mu} 
- \tfrac 12 h^{\lambda \mu} 
+ \tfrac 38 h^{\lambda \tau} {h^\mu}_\tau
+ \cO(h^3)
\big] \ .
\end{align}
Finally, let us record the expansion of the flat components of the spin connection,
which enter the fermion covariant derivatives:
\begin{align}
{e_c}^\tau \omega_{\tau a b}  
= {\delta_c}^\rho {\delta_a}^\mu {\delta_b}^\nu &\big[
- \tfrac 12 \partial_{\mu} h_{\nu\rho}
- \tfrac 12 h_{\nu\tau} \partial^\tau h_{\mu\rho}
+ \tfrac 12 h_{\rho\tau} \partial_{\mu} {h_\nu}^\tau \nn \\
&
+ \tfrac 12 h_{\nu\tau} \partial_{\mu} {h_\rho}^\tau 
+ \tfrac 14 h_{\nu\tau} \partial_\rho {h_\mu}^\tau
- (\mu \leftrightarrow \nu)+ \cO(h^3)
\big] \ .
\end{align}

  \section{Feynman rules} \label{app_Feynman}
 
In this appendix we collect the Feynman rules 
that can be extracted from the massive actions 
\eqref{std_action_spin12}, \eqref{std_action_spin32},
\eqref{std_action_tensors}. The propagators are read
from the free actions where $\vecA_\mu =  0$
and $g_{\mu\nu} = \eta_{\mu\nu}$. 
The interaction vertices are obtained by expanding 
\eqref{std_action_spin12}, \eqref{std_action_spin32},
\eqref{std_action_tensors} up to second order in the metric perturbation
$h_{\mu\nu}$, introduced in \eqref{metric_expansion}.

In all Feynman rules symmetrisation with weight 
one  on
graviton polarisation indices is understood. Moreover,
the momenta of vectors and gravitons are always 
taken to be entering the vertex, 
while the momenta of massive fields flow in the same 
direction as specified by the charge arrow.
 
\subsection{Spin-1/2 fermion}

 \begin{align}
\parbox{3.6 cm}{\includegraphics[width=3.4 cm]{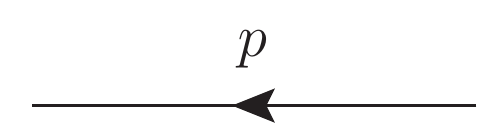}} &= \frac{- \slashed{p} + i c_{1/2} m }{p^2 + m^2} \nn \\
\parbox{4cm}{\includegraphics[width=4cm]{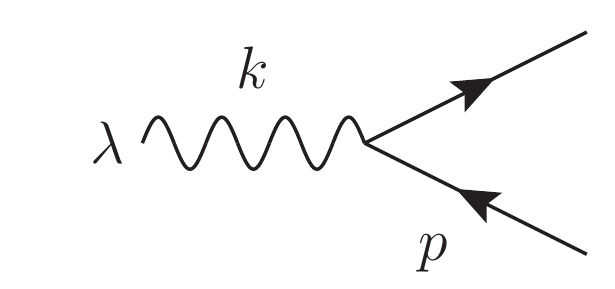}} &= - q \gamma_\lambda \nn \\
\parbox{4cm}{\includegraphics[width=4cm]{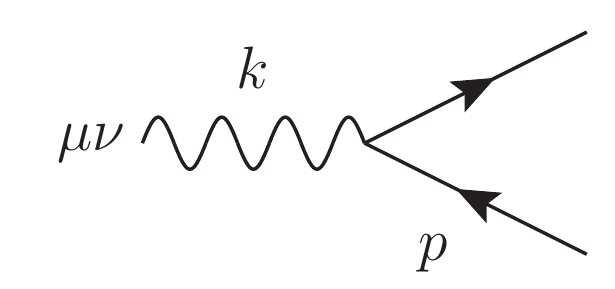}} &=
\tfrac 12 \left( i  c_{1/2} m + \tfrac 12 \slashed{P}   \right) \eta_{\mu\nu} 
- \tfrac 14 \gamma_\mu P_\nu \ , \quad
P \equiv 2p+k \nn 
 \\
\parbox{4cm}{\includegraphics[width=4cm]{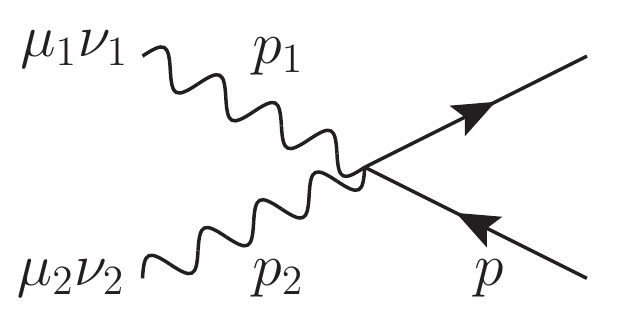}} &= 
\tfrac14 \left( i  c_{1/2} m + \tfrac 12 \slashed{P}   \right) 
\left(   \eta_{\mu_1\nu_1} \eta_{\mu_2 \nu_2} 
- 2 \eta_{\mu_1\mu_2} \eta_{\nu_1 \nu_2}
\right) \nn \\
&+\left[ 
\tfrac{1}{16} \gamma_{\mu_1 \mu_2 \lambda} p_1^\lambda
- \tfrac 18 \gamma_{\mu_1} P_{\nu_1} \eta_{\mu_2 \nu_2}
+ \tfrac{3}{16} \gamma_{\mu_1} P_{\mu_2} \eta_{\nu_1 \nu_2}
+(1 \leftrightarrow 2)
\right] \ , \nn \\
& P \equiv 2p + p_1 +p_2
\nn \\
\parbox{4cm}{\includegraphics[width=4cm]{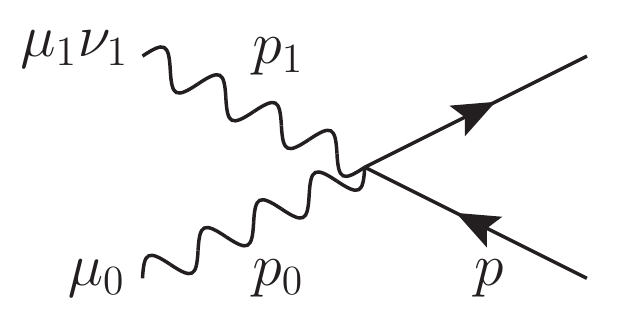}} &= 
 - \tfrac 12 q \eta_{\mu_1 \nu_1} \gamma_{\mu_0} + \tfrac 12 q \eta_{\mu_0 \mu_1} \gamma_{\nu_1} \nn
\end{align}

 \subsection{Massive self-dual tensor}
 
In the following expression antisymmetrisation with weight one on
tensor polarisation indices is understood. 
 \begin{align}
\parbox{4cm}{\includegraphics[width=4cm]{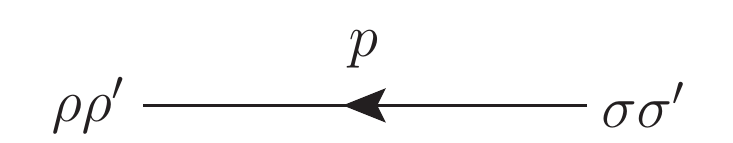}} &=  \frac{1}{p^2 + m^2}
 \left\{
 - i c_B \epsilon_{\rho\rho' \sigma \sigma' \lambda} p^\lambda
 - 2 i m \eta_{\rho\sigma} \eta_{\rho'\sigma'}
 - 4 i m^{-1} \eta_{\rho\sigma} p_{\rho'} p_{\sigma'}
 \right\} \nn \\
\parbox{4cm}{\includegraphics[width=4cm]{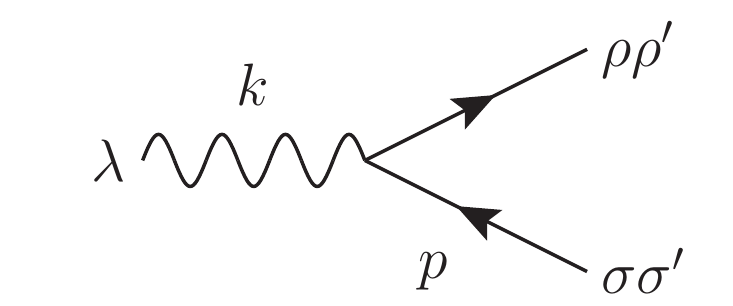}} &=  - \tfrac 14 i c_B q \epsilon_{\rho\rho' \sigma\sigma' \lambda} \nn \\
\parbox{4cm}{\includegraphics[width=4cm]{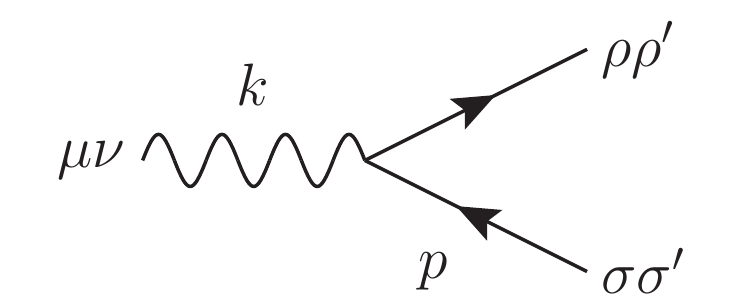}} &=  -\tfrac 14 i m \eta_{\mu\nu} \eta_{\rho\sigma} \eta_{\rho'\sigma'}
 + i m \eta_{\mu \rho} \eta_{\nu \sigma} \eta_{\rho'\sigma'}
 \nn \\ 
\parbox{4cm}{\includegraphics[width=4cm]{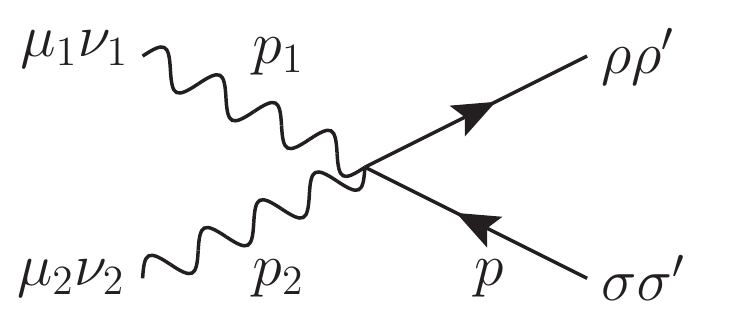}} &=  
  -  \tfrac 18 i m ( \eta_{\mu_1 \nu_1}  \eta_{\mu_2 \nu_2} - 2 \eta_{\mu_1 \mu_2} \eta_{\nu_1 \nu_2} ) \eta_{\rho\sigma} \eta_{\rho' \sigma'}  - i m \eta_{\mu_1 \rho} \eta_{\nu_1 \sigma} \eta_{\mu_2 \rho'} \eta_{\nu_2 \sigma'} \nn \\
 &
 + \left[ 
 \tfrac 1 2 i m \eta_{\mu_1 \nu_1} \eta_{\mu_2 \rho} \eta_{\nu_2 \sigma} \eta_{\rho' \sigma'}
 - i m \eta_{\mu_1 \mu_2} \eta_{\nu_1 \rho} \eta_{\nu_2 \sigma} \eta_{\rho' \sigma'}
 + (1 \leftrightarrow 2)
 \right] \nn \\
\parbox{4cm}{\includegraphics[width=4cm]{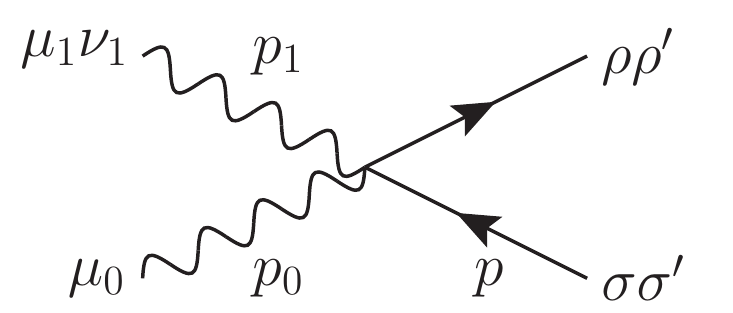}} &= 0 \nn
\end{align}

 \subsection{Spin-3/2 fermion}

\begin{align}
\parbox{4cm}{\includegraphics[width=4cm]{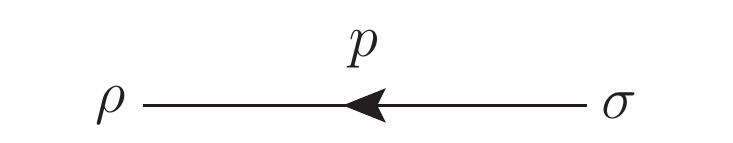}} &= \frac{1}{p^2 + m^2} \, \bigg \{ 
 \left( \eta_{\rho\sigma} + \frac{p_\rho p_\sigma}{m^2} \right)(- \slashed p + i c_{3/2} m) \nn \\
 &+ \frac 14 \left(\gamma_\rho + \frac{i p_\rho}{c_{3/2} m} \right)(- \slashed p - i c_{3/2} m) 
 \left(\gamma_\sigma +  \frac{i p_\sigma}{c_{3/2} m} \right)
 \bigg \} \nn \\
\parbox{4cm}{\includegraphics[width=4cm]{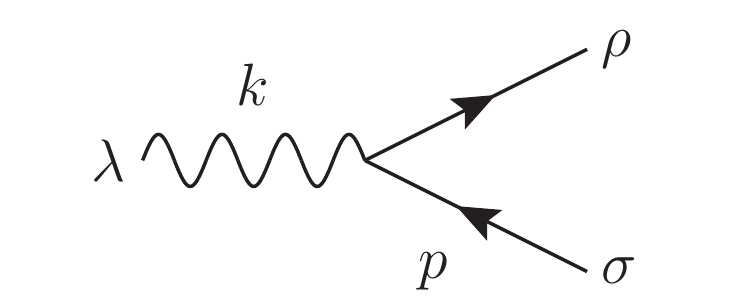}} &= - q \gamma_{\rho \lambda \sigma}  \nn \\
\parbox{4cm}{\includegraphics[width=4cm]{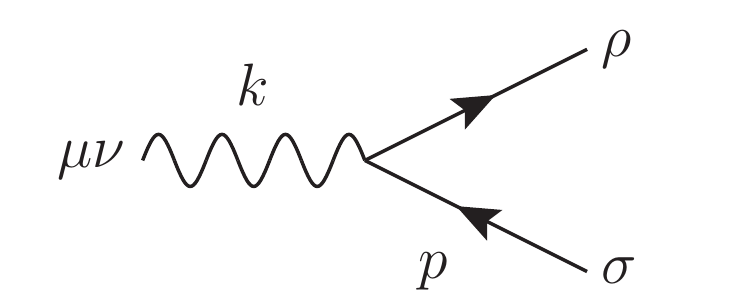}} &= 
\tfrac 12 \left(   \tfrac 12 \gamma_{\rho \lambda \sigma } P^\lambda
- i c_{3/2} m \gamma_{\rho \sigma}
  \right) \eta_{\mu\nu} 
  + \tfrac 14 \gamma_{\rho \sigma \mu} P_\nu \nn \\
 & - \tfrac 14 \eta_{\mu\nu} \gamma_\rho k_\sigma + \tfrac 14 \eta_{\mu\nu} \gamma_\sigma k_\rho
 + \tfrac 14 \eta_{\mu\rho} \gamma_\nu k_\sigma - \tfrac 14 \eta_{\mu\sigma} \gamma_\nu k_\rho
 - \tfrac 14 \eta_{\mu\rho} \gamma_\sigma k_\nu + \tfrac 14 \eta_{\mu\sigma} \gamma_\rho k_\nu \nn \\
 & + \tfrac 12 \left( \tfrac 12 \gamma_{\sigma \lambda \mu} P^\lambda - i c_{3/2} m  \gamma_{\sigma \mu}\right) \eta_{\nu \rho}
 - \tfrac 12 \left( \tfrac 12 \gamma_{\rho \lambda \mu} P^\lambda - i c_{3/2} m  \gamma_{\rho \mu}\right) \eta_{\nu \sigma} \ ,\nn \\
 & P \equiv 2p+k
\nn \\
\parbox{4cm}{\includegraphics[width=4cm]{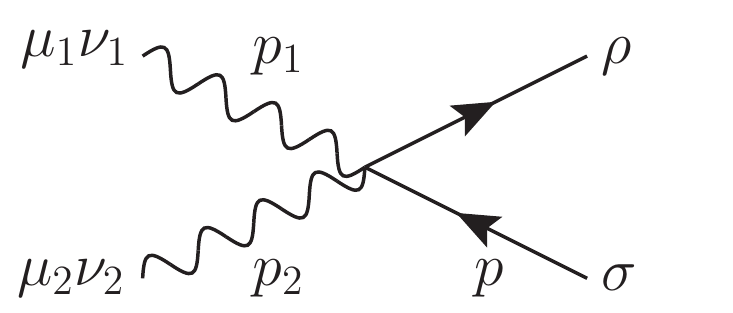}} &= 
\tfrac14 \left( \tfrac 12 \gamma_{\rho \lambda \sigma} P^\lambda   
- i c_{3/2} m \gamma_{\rho\sigma}  \right) 
\left( \eta_{\mu_1 \nu_1} \eta_{\mu_2 \nu_2} - 2 \eta_{\mu_1 \mu_2} \eta_{\nu_1 \nu_2} \right) \nn \\
&+ \big[  - \tfrac 18 \gamma_\rho (p_1 + p_2)_\sigma 
\left( \eta_{\mu_1 \nu_1} \eta_{\mu_2 \nu_2} - 2 \eta_{\mu_1 \mu_2} \eta_{\nu_1 \nu_2} \right) \nn \\
& + \tfrac 38 i c_{3/2} m \gamma_{\sigma \mu_1} \, \eta_{\mu_2 \rho} \, \eta_{\nu_1 \nu_2}
+ \tfrac 14 i c_{3/2} m \gamma_{\mu_1 \mu_2} \, \eta_{\nu_1 \sigma} \, \eta_{\nu_2 \rho} \nn \\
& - \tfrac 18 \gamma_{\sigma \mu_1 \mu_2}\, P_{\nu_1} \, \eta_{\nu_2 \rho} 
+ \tfrac 18 \gamma_{\sigma \mu_1 \mu_2} \, P_{\nu_2} \, \eta_{\nu_1 \rho} 
+ \tfrac 18 \gamma_{\mu_1 \mu_2 \lambda} \, P^\lambda \, \eta_{\nu_1 \rho} \, \eta_{\nu_2 \sigma}
- (\rho \leftrightarrow \sigma) \big]
\nn \\
& + \big[
- \tfrac{1}{16} \gamma_{\rho\sigma \mu_1 \mu_2 \lambda} \, p_1^\lambda \, \eta_{\nu_1 \nu_2}
+ \tfrac 18 \gamma_{\rho \sigma \mu_1}\, P_{\nu_1} \, \eta_{\mu_2 \nu_2}
- \tfrac {3}{16} \gamma_{\rho \sigma \mu_1 }\, P_{\mu_2} \, \eta_{\nu_1 \nu_2}
+ (1 \leftrightarrow 2) \big] \nn \\
&+ \big[ - \tfrac 18   \gamma_\sigma \, p_{1\, \mu_1}\,  \eta_{\nu_1 \rho} \, \eta_{\mu_2 \nu_2}
- \tfrac {1}{16} \slashed{p}_1 \, \eta_{\mu_1 \mu_2} \, \eta_{\nu_1 \sigma} \eta_{\nu_2 \rho}
+ \tfrac{3}{16} \gamma_\sigma \, p_{1\, \mu_2} \, \eta_{\mu_1 \rho} \, \eta_{\nu_1 \nu_2} \nn \\
& + \tfrac{5}{16} \gamma_\sigma p_{1\, \mu_1} \, \eta_{\mu_2 \rho} \, \eta_{\nu_1 \nu_2}
- \tfrac 18 \gamma_{\mu_2} p_{1\, \mu_1} \, \eta_{\nu_1 \sigma} \, \eta_{\nu_2 \rho}
- \tfrac 14 \gamma_\sigma p_{1\, \mu_2} \, \eta_{\nu_2 \rho} \, \eta_{\mu_1 \nu_1} \nn \\
& + \tfrac 14 \gamma_{\mu_1} p_{1\, \mu_2} \, \eta_{\nu_2 \rho} \, \eta_{\nu_1 \sigma}
+ \tfrac 18 \gamma_{\mu_2} p_{1\, \rho} \, \eta_{\mu_1 \sigma} \, \eta_{\nu_1 \nu_2}
- \tfrac 18 \gamma_{\mu_1} p_{1\, \rho} \, \eta_{\nu_1 \sigma} \, \eta_{\mu_2 \nu_2} \nn \\
& - \tfrac 18 \gamma_{\mu_2} p_{1\, \rho} \, \eta_{\mu_2 \sigma} \, \eta_{\mu_1 \nu_1}
+ \tfrac 14 \gamma_{\mu_1} p_{1\, \rho} \, \eta_{\mu_2 \sigma} \, \eta_{\nu_1 \nu_2} 
 - \tfrac 14 i c_{3/2 } m \gamma_{\sigma \mu_1} \, \eta_{\nu_1 \rho} \, \eta_{\mu_2 \nu_2} \nn \\
 & + \tfrac{3}{16} \gamma_{\sigma \mu_2 \lambda} \, P^\lambda \, \eta_{\mu_1 \rho} \, \eta_{\nu_1 \nu_2}
 - \tfrac 18 \gamma_{\sigma \mu_1 \lambda} \, P^\lambda \, \eta_{\nu_1 \rho} \, \eta_{\mu_2 \nu_2} 
-(\rho \leftrightarrow \sigma )
+ (1 \leftrightarrow 2)
\big] \ , \nn \\
& P \equiv 2 p + p_1 + p_2
 \nn \\
\parbox{4cm}{\includegraphics[width=4cm]{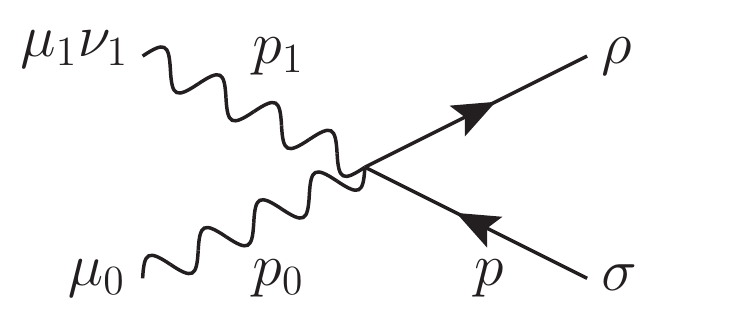}} &= 
- \tfrac 12 q \eta_{\mu_0 \mu_1} \gamma_{\nu_1 \rho\sigma}
+ \tfrac 12 q \eta_{\mu_1 \nu_1} \gamma_{\mu_0 \rho\sigma}
- \tfrac 12 q \eta_{\mu_1 \rho} \gamma_{\mu_0 \nu_1 \sigma}
+ \tfrac 12 q \eta_{\mu_1 \sigma} \gamma_{\mu_0 \nu_1 \rho} \nn
\end{align}


 \section{Torus integration}\label{App:Torus}
In this appendix we perform the integral (\ref{LatticeIntegral}) over the world-sheet 
torus. It can be done in a straight-forward manner using lattice reduction 
techniques (see \cite{Borcherds}).\footnote{Other techniques for computing such integrals can be found in \cite{Dixon:1990pc}.} To this end, we begin by introducing the 
Siegel-Narain theta-function (without insertions) of the $\Gamma^{(1,1)}$ Narain lattice
\begin{align}
\Theta^{(1,1)}(\tau,\bar{\tau};P):=\sum_{\lambda\in\Gamma^{(1,1)}}e^{\pi i\tau (P_+(\lambda))^2-\pi i\bar{\tau} (P_-(\lambda))^2}\,,
\end{align}
and pick a basis of lattice vectors $\Gamma^{(1,1)}=\langle e,f\rangle_{\mathbb{Z}}$. 
Moreover, we view the Narain momenta introduced above as projections $P_\pm$ of 
the isometry $P:\Gamma\otimes \mathbb{R}\rightarrow\mathbb{R}^{1,1}$ to $\mathbb{R}^{1,0}$ and $\mathbb{R}^{0,1}$ respectively
\begin{align}
&\begin{array}{l}P_+(\lambda):=P_L=\frac{1}{\sqrt{2}}\left(\frac{m}{R}-nR\right)\,,\\
P_-(\lambda):=P_R=\frac{1}{\sqrt{2}}\left(\frac{m}{R}+nR\right)\,,\end{array}&&\text{for} &&\lambda=(n;m)\in\Gamma^{(1,1)}\,,\label{NarainMomenta}
\end{align}
where we have used the same parameterisation as in \cite{Antoniadis:1995vz} in terms 
of the modulus $\rad$. Following the notation of \cite{Marino:1998pg}, we can pick a 
primitive null vector $z\in\Gamma^{(1,1)}$ along which we can reduce $\Gamma^{(1,1)}$
to the trivial lattice. Upon denoting $z_\pm:=P_\pm(z)$ we get for (\ref{LatticeIntegral}) \cite{Borcherds}
\begin{align}
\mathcal{I}(\rad)=\frac{\mathcal{I}_{\text{red}}}{\sqrt{2z_+^2}}+\sqrt{\frac{2}{z_+^2}}\sum_{\ell=1}^\infty\sum_{t=0}^1c(0,t)\left(\frac{2z_+^2}{\pi\ell^2}\right)^{t+1}\Gamma(t+1)\,.\label{ResultTorusIntegral}
\end{align}
Here $\mathcal{I}_{\text{red}}$ is another torus integral, however, without a 
Siegel-Narain theta function in the integrand
\begin{align}
\mathcal{I}_{\text{red}}=\frac{1}{2\zeta(2)}\int \frac{d^2\tau}{\tau_2^2}\,\hat{\bar{G}}_2(\tau,\bar{\tau})\bar{F}(\bar{\tau})=\frac{3}{2\pi^3}\left[\bar{G}_2^2(\bar{\tau})\,\bar{F}(\bar{\tau})\right]\bigg|_{\bar{q}^0}=-48\pi\,d_0\,.
\end{align}
with $G_{2k}:=2\zeta(2k)E_{2k}$. Moreover, the information about the modular form $\bar{F}(\bar{\tau})$ enters into (\ref{ResultTorusIntegral}) through the coefficients $c(0,t)$. To be precise, they are the Fourier coefficients of the expansion of the following modular invariant function
\begin{align}
\hat{E}_2(\bar{\tau})F(\bar{\tau})=\sum_{k=-1}^\infty\sum_{t=0}^1c(k,t)\,\bar{q}^{k}\tau_2^{-t}\,.
\end{align}
Before giving the explicit expression for $\mathcal{I}(\rad)$, there is one more 
crucial point which needs mentioning. The expression (\ref{ResultTorusIntegral}) 
is only valid under the assumption that $z_+$ is sufficiently small. Since $z_+$ 
(through the projection $P_+$) depends on the modulus $\rad$, a particular choice 
of $z$ will only be valid in a certain region of the moduli space (chamber dependence). 
Therefore, in order to cover all chambers, we need to consider reductions along different 
vectors. A convenient choice is to set $z=e$, in which case (see \cite{Marino:1998pg}) 
$z_+^2=\frac{\rad^2}{2}$ which is a valid choice for $\rad<1$. 
From this we can deduce the result in the chamber $\rad>1$ by 
exchanging $\rad\leftrightarrow \tfrac{1}{\rad}$. The full answer then takes the form
\begin{align}
\mathcal{I}(\rad)&=-d_0\pi\,\theta(1-\rad)\,\left[\frac{48}{\rad}-2\sum_{t=0}^1\frac{\zeta(2+2t)\Gamma(t+1)}{\pi^{t+2}}\,\frac{c(0,t)}{d_0}\,\rad^{2t+1}\right]+\left(\rad\leftrightarrow\tfrac{1}{\rad}\right)\nonumber\\
&=-8d_0\pi\,\left[6\left(\rad+\frac{1}{\rad}\right)+5\left(\rad\theta(1-\rad)+\frac{\theta(\rad-1)}{\rad}\right)-2\left(\rad^3\theta(1-\rad)+\frac{\theta(\rad-1)}{\rad^3}\right)\right]\,,
\end{align}
This expression is also consistent with the expression 
for $\mathcal{I}_0$ in (\ref{AFTintegral}) first computed in \cite{Antoniadis:1995vz}.



\end{document}